\documentclass[useAMS,usenatbib]{mn2e}
\usepackage{epsfig,longtable}

\def\spose#1{\hbox to 0pt{#1\hss}}
\def\lta{\mathrel{\spose{\lower 3pt\hbox{$\mathchar"218$}}
     \raise 2.0pt\hbox{$\mathchar"13C$}}}
\def\gta{\mathrel{\spose{\lower 3pt\hbox{$\mathchar"218$}}
     \raise 2.0pt\hbox{$\mathchar"13E$}}}
 
\title[]{The cosmic evolution of radio-AGN feedback to ${\mathbf z = 1}$}

\author[P.~N.~Best et~al.]{P.~N.~Best$^1$\thanks{Email:
    pnb@roe.ac.uk}, L.~M.~Ker$^1$, C. Simpson$^2$, E.~E.~Rigby$^3$, J. Sabater$^1$
\\
$^1$ SUPA\thanks{Scottish Universities Physics Alliance}, Institute for
Astronomy, Royal Observatory Edinburgh, Blackford Hill, Edinburgh EH9 3HJ \\
$^2$ Astrophysics Research Institute, Liverpool John Moores University, Liverpool Science
Park, 146 Brownlow Hill, Liverpool L3 5RF\\
$^3$ Sterrewacht Leiden, Postbus 9513, 2300 RA Leiden, the Netherlands\vspace*{-2mm}}

\begin{document}

\pagerange{\pageref{firstpage}--\pageref{lastpage}}
\pubyear{2013}

\label{firstpage}

\maketitle

\begin{abstract}
\noindent 

This paper presents the first measurement of the radio luminosity function
of `jet-mode' (radiatively-inefficient) radio-AGN out to $z=1$, in order
to investigate the cosmic evolution of radio-AGN feedback.  Eight radio
source samples are combined to produce a catalogue of 211 radio-loud AGN
with $0.5<z<1.0$, which are spectroscopically classified into jet-mode and
radiative-mode (radiatively-efficient) AGN classes. Comparing with large
samples of local radio-AGN from the Sloan Digital Sky Survey, the cosmic
evolution of the radio luminosity function of each radio-AGN class is
independently derived. Radiative-mode radio-AGN show an order of magnitude
increase in space density out to $z \approx 1$ at all luminosities,
consistent with these AGN being fuelled by cold gas. In contrast, the
space density of jet-mode radio-AGN decreases with increasing redshift at
low radio luminosities ($L_{\rm 1.4 GHz} \lta 10^{24}$W\,Hz$^{-1}$) but
increases at higher radio luminosities. Simple models are developed to
explain the observed evolution. In the best-fitting models, the
characteristic space density of jet-mode AGN declines with redshift in
accordance with the declining space density of massive quiescent galaxies,
which fuel them via cooling of gas in their hot haloes. A time delay of
1.5--2\,Gyr may be present between the quenching of star formation and the
onset of jet-mode radio-AGN activity. The behaviour at higher radio
luminosities can be explained either by an increasing characteristic
luminosity of jet-mode radio-AGN activity with redshift (roughly as
$(1+z)^3$) or if the jet-mode radio-AGN population also includes some
contribution of cold-gas-fuelled sources seen at a time when their
accretion rate was low. Higher redshifts measurements would distinguish
between these possibilities.
\end{abstract}

\begin{keywords}
galaxies: active --- radio continuum: galaxies --- galaxies: jets ---
accretion, accretion discs --- galaxies: evolution
\end{keywords}

\section{Introduction}
\label{sec_intro}

Understanding the evolution of galaxies, from the end of the `dark ages'
through to the complexity and variety of systems we observe in the local
Universe, remains a primary goal for observational and theoretical
astrophysics. A crucial piece in the picture is the role that active
galactic nuclei (AGN) play in controlling or terminating the star
formation of their host galaxies \citep[see reviews
  by][]{cat09a,fab12,hec14}. Over recent years it has become clear that
AGN activity falls into two fundamental modes, each of which may have a
distinct `feedback' role in galaxy evolution. Accretion at high fractions
($\gta 1$\%) of the Eddington rate produces radiatively-efficient
(quasar/Seyfert-like; hereafter `{\it radiative-mode}') AGN, which display
luminous radiation from a geometrically thin, optically thick accretion
disk \citep[e.g.][]{sha73}. Accretion at low Eddington fractions leads to
an advection-dominated accretion flow (e.g.\ Narayan \& Yi 1994,
1995)\nocite{nar94,nar95}; these AGN (hereafter `{\it jet-mode}' AGN) are
radiatively inefficient, and the bulk of their energetic output is in
kinetic form, in two-sided collimated outflows (jets). For a full review of these
two AGN populations, and their host galaxy properties, the reader is
referred to \citet{hec14}.

The role of radiative-mode AGN in galaxy evolution remains hotly debated.
These AGN are frequently invoked to quench star formation in massive
galaxies, causing these to migrate from the locus of star-forming galaxies
on to the red sequence \citep[e.g.][]{sil98,hop05,spr05c,sch07,cim13}.
Models of this process can provide an explanation for the relationship
seen between black hole mass and bulge velocity dispersion
\citep{sil98,fab99,kin03}.  However, although there is ample evidence that
radiative-mode AGN can drive winds \citep[e.g.\ see reviews
  by][]{vei05,fab12}, observational evidence for galaxy-scale feedback
from radiative-mode AGN is so far limited to only the extreme, high
luminosity systems, with little evidence that it occurs in more typical
systems. Most radiative-mode AGN appear to be associated with star-forming
galaxies, and to be fuelled by secular processes \citep[][and references
  therein]{hec14}. There are also indications that secular processes,
rather than AGN activity, could be responsible for the quenching of
star formation \citep[`mass-quenching';][]{pen10,sch14} and possibly
setting up the black hole mass relations \citep[e.g.][]{lar10,jah11}.
There remains much to be understood about whether radiative-mode AGN play
any significant role in galaxy evolution.

In contrast, it is now widely accepted that recurrent jet-mode AGN
activity is a fundamental component of the lifecycle of the most massive
galaxies, responsible for maintaining these galaxies as `red and dead'
once they have migrated on to the red sequence
\citep[e.g.][]{cro06,bow06,bes06a,fab06}. This is achieved by the radio
jet depositing the AGN energy in kinetic form into the local intergalactic
medium, through bubbles and cavities inflated in the surrounding hot gas
\citep{boh93,car94b,mcn00,fab03}. This energy counteracts the radiative
energy losses of that hot gas and prevents the bulk of the gas from
cooling. This is most readily observed in the central galaxies of
cool-core clusters (those with cooling times well below a Gyr, for which a
counter-balancing heating source is required): in these systems, both
radio AGN activity \citep{bur90,bes07} and X-ray cavities
\citep{dun06,fab12} are almost universally present, and the current jet
mechanical luminosity is seen to balance the cooling luminosity \citep[see
  review by][]{mcn07}. The jet-mode AGN are believed to be fuelled
primarily by the cooling of hot gas in the interstellar and intergalactic
medium, and they deposit their energy back into this same hot gas,
providing the necessary conditions for a feedback cycle \citep[][and
  references therein]{hec14}.

Amongst the general massive galaxy population, the prevalence of jet-mode
AGN is a very strong function of both stellar mass ($M_*$) and black hole
mass ($M_{\rm BH}$), with the fraction of galaxies hosting a jet-mode
radio-AGN scaling as $M_*^{2.5}$ and as $M_{\rm BH}^{1.6}$
\citep{bes05b,jan12}. In these systems, the instantaneous mechanical
luminosity of the AGN jets can greatly exceed the cooling luminosity of
the hot gas surrounding the galaxy, but if account is taken of the duty
cycle of the recurrent activity then the time-averaged jet mechanical
energy output is in closer agreement with the cooling losses
\citep{bes06a}.  The jet-mode AGN appear to act as a cosmic thermostat,
being switched on whenever the cooling rate of the hot gas rises above
some threshold, and acting to inhibit the gas cooling (and therefore
switch off the AGN's own gas supply as well). In the most massive
galaxies, after the AGN switches off, the cooling quickly recommences, and
so the AGN duty cycle is short and the prevalence of jet-mode AGN is
high. In less massive galaxies, the AGN remains switched off for a longer
time since the lower binding energy and gas sound speed lead to a longer
recovery time before gas cooling and accretion recommence
\citep[e.g.][]{gas13}: these systems have a lower AGN prevalence, and
oscillate around an equilibrium state.

This picture of jet-mode AGN activity has been established through
detailed studies of the nearby Universe, and an important test of its
validity is to examine whether it is consistent with observations at
earlier cosmic times. An easily testable prediction of the model is that
if jet-mode AGN are fuelled in the same manner at all redshifts, then the
steep relationship between AGN prevalence and stellar mass ought to remain
in place at higher redshift. Early studies of this, out to $z \lta 1$,
indicate that the same relation is indeed seen
\citep{tas08,don09,sim13}. A second measurable property is the cosmic
evolution of the space density of jet-mode AGN. Phenomenological models of
the dual-populations of AGN predict that the space density of jet-mode AGN
activity should remain roughly flat out to moderate redshifts \citep[$z
  \sim 1$;][]{cro06,mer08,kor08,moc13}, but observationally this remains
unconstrained.  Measuring this is the focus of the current paper.

By far the best way to trace the cosmic evolution of the jet-mode AGN is
through radio-selected samples, directly tracing the radio jet
activity. The evolution of the radio luminosity function (RLF) of
radio-loud AGN has been well-studied over many decades: it is known to be
strongly luminosity dependent with the most powerful sources showing very
rapid cosmic evolution \citep[a factor $\sim$thousand increase in space
  density out to redshift 2--3; cf.][and references therein]{dun90,rig11},
while less powerful sources show only a modest (factor 1.5--2) space
density increase out to $z \sim 0.5$ \citep[e.g.][]{sad07,don09} with a
possible decline thereafter \citep{rig11,sim12}. However, the RLF is
composed not only of jet-mode AGN, but also of the population of
radio-loud radiative-mode AGN: these comprise the radio-loud quasars and
their edge-on counterparts (often referred to as `High-Excitation Radio
Galaxies'). In order to observationally determine the cosmic evolution of
just the jet-mode AGN (`Low-Excitation Radio Galaxies'), it is necessary
to separate these two contributions to the overall RLF of radio-AGN.

Radiative-mode AGN dominate the radio-AGN population at higher radio
luminosities where strong cosmic evolution is seen, while jet-mode
radio-AGN dominate the radio population at lower radio luminosities, where
cosmic evolution is far weaker. This has led many authors to consider a
simple division in radio luminosity to separate the two radio
populations. However, \citet[][hereafter BH12]{bes12} used data from the
Sloan Digital Sky Survey \citep[SDSS;][]{yor00,str02} to classify a local
population of radio-AGN, and showed that both radiative-mode and jet-mode
radio-AGN are found across all radio luminosities. They also provided
evidence that, at a given radio luminosity, the two AGN classes show
distinct cosmic evolution. This indicates that explicit separation of the
two radio populations is needed to directly determine the cosmic evolution
of jet-mode AGN alone.

This paper assembles and spectroscopically classifies a large sample of
radio-AGN with $0.5 < z < 1.0$ across a broad range of radio luminosity,
by combining eight radio surveys from the literature with high
spectroscopic completeness, and adding additional spectroscopic
observations. The samples are presented in Section~\ref{sec_samples},
where the local comparison sample is also defined. Classification of the
sources is described in Section~\ref{sec_classify}. In
Section~\ref{sec_rlfs}, these data are used to determine the cosmic
evolution of the RLF of jet-mode AGN, and simple models are developed to
explain the observed evolution. The implications of the results are
discussed in Section~\ref{sec_discuss}, and conclusions are drawn in
Section~\ref{sec_concs}. Throughout the paper, the cosmological parameters
are assumed to have values of $\Omega_m = 0.3$, $\Omega_{\Lambda} = 0.7$,
and $H_0 = 70$\,km\,s$^{-1}$Mpc$^{-1}$.\vspace*{-2mm}

\section{Radio Source Samples}
\label{sec_samples}

\subsection{The local radio-AGN populations}
\label{sec_local}

BH12 combined spectroscopic data from the `main galaxy sample' of the SDSS
with radio data from the National Radio Astronomy Observatory (NRAO) Very
Large Array (VLA) Sky Survey \citep[NVSS;][]{con98} and the Faint Images
of the Radio Sky at Twenty centimetres (FIRST) survey \citep{bec95} to
derive a sample of over 7000 radio-loud AGN in the local Universe. Both
star-forming galaxies and radio-quiet quasars were excluded from their
sample. Using the wide range of emission line flux measurements available
for these sources, in conjunction with the line equivalent widths and the
emission line to radio luminosity distributions, BH12 classified the
sources as either jet-mode or radiative-mode radio-AGN.

\begin{figure}
\centerline{
\psfig{file=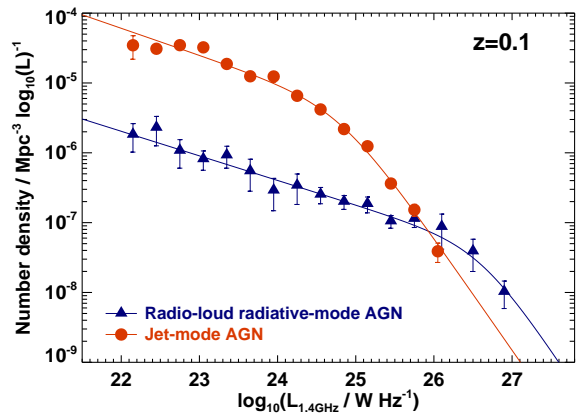,width=8.3cm,clip=} 
}
\caption{\label{fig_localrlfs} The local radio luminosity functions for
  the radiative-mode and jet-mode radio-AGN separately, along with
  best-fitting broken power-law models \citep[adapted from][]{hec14}.}
\end{figure}

BH12 determined the local RLFs for the two AGN classes. However, since
their radio source sample was based upon the SDSS main galaxy sample it
excluded both radio-loud quasars and broad-line radio galaxies; these can
be dominant in the radiative-mode AGN population at higher radio
luminosities.  \citet{gen13} also derived RLFs for jet-mode and
radiative-mode sources, using the Combined NVSS-FIRST Galaxy catalogue
(CoNFIG) which, although much smaller, did not suffer from this bias. They
found broad agreement with BH12 except for the high luminosity
radiative-mode AGN. For the analysis of this paper, therefore, the local
RLFs were constructed using primarily the BH12 results, but replacing
these for radiative-mode AGN above $L_{\rm 1.4 GHz} = 10^{26}$
W\,Hz$^{-1}$ by the steep-spectrum\footnote{Analysis is limited to
  steep-spectrum ($\alpha > 0.5$ where $S_\nu \propto \nu^{-\alpha}$)
  sources to avoid the complications of beamed emission. \citet{gen13} did
  not remove flat-spectrum sources from their RLFs, but their contribution
  is small and removing them changes the space density estimates by less
  than the uncertainties. Spectral indices are not available for most BH12
  sources but flat-spectrum sources are expected to be rare in this
  population. All of the higher redshift samples described in
  Section~\ref{sec_highzsamp} are limited to only steep-spectrum sources.}
RLF determined from the Gendre et~al.\ CoNFIG data
\citep[cf.][]{hec14}. The resultant (steep-spectrum) local RLFs are shown
in Figure~\ref{fig_localrlfs}, along with the best-fitting broken power
law models of the form

\begin{displaymath}
\rho = \frac{\rho_0}{(L/L_0)^{\beta} + (L/L_0)^{\gamma}} 
\end{displaymath}

\noindent where $L_0$ is a characteristic luminosity and $\rho$ and
$\rho_0$ are measured in units of number of sources per log$_{10}L$ per
Mpc$^3$.

The BH12 sample also provides the basis for the optimisation of emission
line ratio diagnostics in Section~\ref{sec_classify} to segregate the
jet-mode and radiative-mode sources at the higher redshifts (where far
fewer emission line fluxes are available).

\subsection{A combined radio source sample at ${\mathbf 0.5 < z <
    1.0}$}
\label{sec_highzsamp}

Eight separate radio surveys with a wide range of flux density limits were
combined in order to develop a large total radio source sample covering a
broad span of radio luminosities. Each of these surveys was selected to
have high spectroscopic completeness from which galaxies in the target
redshift range $0.5 < z < 1.0$ could be drawn. Where necessary, each
sample was re-selected at 1.4\,GHz, as described below, to produce a
sample which would be complete for steep-spectrum sources down to a fixed
1.4\,GHz flux density limit over the same sky area as the original
sample. This re-selection avoids selection biases in the analysis.

Despite the use of the highest spectroscopic completeness samples
available, a significant number of radio sources lacked either
spectroscopic redshifts or available spectra of sufficient quality to
allow classification as jet-mode or radiative-mode. A programme of
spectroscopic follow-up observations was therefore carried out, targeting
objects which either lacked classification, or which had photometric
redshifts within (or very close to) the target range $0.5 < z < 1.0$. This
spectroscopic programme was carried out on the William Herschel Telescope
(WHT) during two runs in 2012; details of the observations and the results
arising are described in Appendix~\ref{app_specdetails}. On the basis of
these new observations, some sources were removed from the samples as
their spectroscopic redshifts placed them outside the required
range\footnote{Likewise, it is undoubtedly the case that some sources,
  lacking spectra, will have been excluded from the sample because their
  photometric redshift places them outside the target redshift range, but
  whose spectroscopic redshift actually lies within the range, meaning
  that they should have been included. However, the number of such sources
  is expected to be sufficiently small that their exclusion will not
  significantly influence any of the results of this paper.}.  Details of
these excluded sources are given in Table~\ref{tab_excluded}. In the
descriptions of the 8 samples that follows, these objects are already
excluded when discussing numbers of sources.

\begin{table}
\caption{\label{tab_surveys} Properties of the radio surveys used for the
  construction of the $0.5 < z < 1.0$ sample.}
\begin{tabular}{cccc}
\hline
Survey      & Sky area & Flux density lim. & No of sources \\
            &  (sr)    & (mJy, 1.4\,GHz) & ($0.5<z<1.0$) \\
\hline
WP85r       &  9.81    &     4000           &      29       \\     
CoNFIG-1    &  1.50    &     1300           &      45       \\
CoNFIG-2r   &  0.89    &      800           &      24       \\ 
PSRr        &  0.075   &      500           &      9       \\
7CRSr       &  0.022   &      167           &      21       \\
TOOT-00r    &  0.0015  &       33           &       7       \\ 
CENSORS     &  0.0018  &      7.2           &      28       \\
Hercules    &  0.00038 &      2.0           &      16       \\
SXDS        &  0.000247&      0.2           &      32       \\
\hline
\end{tabular}
\end{table}

\subsubsection{Wall \& Peacock sample}

The original \citet{wal85} radio source sample contained 233 radio sources
brighter than 2.0\,Jy at 2.7\,GHz over 9.81\,sr of sky. From this,
\citet{rig11} re-selected a sample of 138 steep spectrum ($\alpha > 0.5$)
radio sources which was complete to a flux density limit of 4\,Jy at
1.4\,GHz. This re-selected sample, hereafter referred to as WP85r, is 97\%
spectroscopically complete, and contains 29 sources in the redshift range
$0.5 < z < 1.0$ (including one photometric redshift source).

\subsubsection{CoNFiG sample}

The CoNFIG catalogue was presented by \citet{gen10} and consists of four
different radio source samples selected at 1.4\,GHz from the NVSS at
different flux density levels. Here, the CoNFiG-1 sample is used, along
with the revised `CoNFIG-2r' sample defined by \citet{ker12}. CoNFiG-1 is
complete to a 1.4\,GHz flux density limit of 1.3\,Jy. CoNFiG-2r
corresponds to the subset of CoNFiG-2 with flux densities in the range
0.8\,Jy $< S_{\rm 1.4 GHz} <$ 1.3\,Jy; the lower flux density limit is set
because at fainter flux densities CoNFIG-2 becomes rapidly more incomplete
in terms of optical identifications and redshift estimates, while sources
brighter than 1.3\,Jy are already in CoNFIG-1 since the sky areas overlap.
The combined CoNFIG-1 and CoNFIG-2r samples contain 6 steep spectrum
sources without optical identification, but the magnitude limits indicate
that these have redshifts $z \gta 1$ \citep[see discussion in][]{ker12},
so these sources are discounted for the current analysis. Excluding also
three further sources which are duplicates of WP85r sources (3C196, 3C237,
3C280), there are 45 CoNFIG-1 and 24 CoNFIG-2r sources with spectroscopic
(60) or photometric (9) redshifts in the range $0.5 < z < 1.0$.

\subsubsection{Parkes Selected Regions sample}

The original Parkes Selected Regions sample \citep{wal71,dow86,dun89b} was
defined at 2.7\,GHz and contains 178 radio sources brighter than 0.1 Jy
over 0.075\,sr of sky. \citet{rig11} re-selected the sample at 1.4\,GHz to
produce a complete sample of 59 steep spectrum ($\alpha > 0.5$) sources
above a flux density of $S_{\rm 1.4 GHz} = 0.36$\,Jy. 20 of these sources
have redshifts in the redshift range $0.5 < z < 1.0$. However, at the
faintest flux densities the spectroscopic classification fraction is low,
so the sample used here (referred to as PSRr) is restricted to the 9
sources above $S_{\rm 1.4 GHz} = 0.50$\,Jy.

\subsubsection{7C Redshift Survey sample}

The Seventh Cambridge Redshift Survey, 7CRS, is composed of three
subsamples, 7CI, 7CII and 7CIII, over three different sky areas totalling
0.022\,sr, each selected at 151\,MHz down to a limiting flux density limit
of around 0.5\,Jy \citep[][and references therein]{wil02a,lac99b}. This
sample was re-selected at 1.4\,GHz down to a flux density of $S_{\rm 1.4
  GHz} = 0.167$\,Jy. Although this re-selection removes a large fraction
of the 7CRS sample, the remaining sample will be complete for steep
spectrum ($\alpha > 0.5$) sources, and populates an otherwise
sparsely-sampled range of radio luminosities. 21 sources from the
re-selected sample, hereafter referred to as 7CRSr, have redshifts (all
spectroscopic) in the target range.

\subsubsection{Tex-Ox One Thousand sample}

The Tex-Ox One Thousand (TOOT) survey \citep{hil03} was an ambitious
attempt to measure spectroscopic redshifts for 1000 galaxies down to
$S_{\rm 151 MHz} = 0.1$\,Jy; so far only results in the TOOT-00 field have
been published \citep{var10}. The sample, over a sky area of 0.0015\,sr,
has been re-selected at 1.4\,GHz down to a flux density limit of
0.033\,Jy, above which it will be complete for steep spectrum sources. 7
sources in this re-selected (TOOT-00r) sample lie between redshifts 0.5
and 1.0 (all spectroscopically confirmed).

\subsubsection{CENSORS sample}

The Combined EIS-NVSS Survey of Radio Sources (CENSORS) is a
1.4\,GHz-selected sample of 135 radio sources down to a flux density limit
of 0.0072\,Jy, over the 0.0018\,sr sky region that overlaps the ESO
Imaging Survey (EIS) patch D \citep{bes03b,bro06,bro08,rig11}. At nearly
80\% spectroscopically complete, it is one of the most complete faint
radio source samples available. 28 of these sources lie in the redshift
range $0.5 < z < 1.0$ (including four photometric redshifts).

\subsubsection{Hercules sample}

The Hercules sample is taken from a field in the Leiden-Berkeley Deep
Survey \citep{win84}, and consists of 64 sources selected to have a flux
density greater than 0.002\,Jy at 1.4 GHz \citep{wad01}. The surveyed sky
area is 0.00038\,sr. 16 of these sources have spectroscopic (14) or
photometric (2) redshifts between 0.5 and 1.0.

\subsubsection{SXDF sample}

A deep 1.4\,GHz radio survey of the Subaru/XMM-Newton Deep Field, which
overlaps the United Kingdom Infrared Deep Sky Survey \citep[UKIDSS;
][]{law07} Ultra-Deep Survey (UDS) region, has been carried out by
\citet{sim06}. These data reach a depth of 12$\mu$Jy rms in the central
regions, with the catalogue complete to the 100$\mu$Jy level (for point
sources) over the whole field. Spectroscopic and photometric redshift data
for the detected radio sources were presented by \citet{sim12}. The
spectroscopic completeness decreases at lower flux densities, so to reduce
the number of unclassified sources, the current analysis was restricted to
sources with an integrated flux density level above $S_{\rm 1.4 GHz} =
0.0002$\,Jy, over the survey area of 0.000247\,sr. This flux density cut
also reduces the risk of missing faint extended sources. Within the
redshift range $0.5 < z < 1.0$, 38 sources were selected.

At the depth of this survey, starbursting galaxies and radio-quiet quasars
(both optically obscured and unobscured) are expected to contribute a
significant fraction of the radio source population (their contribution to
other brighter samples is expected to be negligible). \citet{sim12}
identified starburst galaxies within the sample from their emission line
ratios and absorption line properties. Radio-quiet quasars can be
identified on the basis of the ratio between their mid-infrared 24$\mu$m
flux density and their radio flux density (the $q_{24}$ parameter, where
$q_{24} = {\rm log}_{10}(S_{24\mu m} / S_{\rm 1.4GHz})$ and $S_{24\mu m}$ and
$S_{\rm 1.4GHz}$ are each k-corrected values). Star-forming galaxies and
radio-quiet quasars both display a narrow distribution in this parameter
\citep[e.g.][]{app04,iba08,sim12}, while radio-loud AGN are offset to
lower values.  $q_{24}$ values for the SXDF were calculated by
\citet{sim12} and the threshold value of $q_{24} = -0.23$ determined by
\citet{iba08} was adopted to remove objects with higher $q_{24}$
values. In this manner, a clean sample of 27 radio-loud AGN was selected
in the target redshift range (including 5 sources with photometric
redshifts).

\subsubsection{Summary of combined sample}

\begin{figure}
\centerline{
\psfig{file=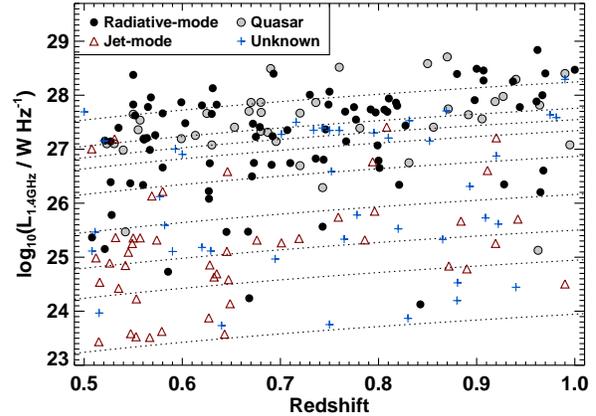,width=8.6cm,clip=} 
}
\caption{\label{fig_pzdist} The distribution of the combined radio sample
  on the radio luminosity versus redshift plane. Sources are classified as
  radiative-mode and jet-mode where possible. Sources classified as
  quasars are plotted with separate symbols but form part of the
  radiative-mode class. The dotted lines indicate the luminosity limits of
  the different samples as a function of redshift for a spectral index
  $\alpha = 0.75$.}
\end{figure}

The combined $0.5<z<1.0$ sample contains 211 radio sources (including 27
with photometric redshifts). The properties of these sources are provided
in Table~\ref{tab_sample} and their distribution on the radio luminosity
versus redshift plane is shown in Figure~\ref{fig_pzdist}.

\section{Classification of the radio sources} 
\label{sec_classify}

The classification of radio galaxies as radiative-mode or jet-mode can be
carried out using emission line strengths and line flux ratios.  In the
nearby Universe, the high quality of the SDSS spectroscopic data allowed
BH12 to derive reliable classifications using the emission line ratio
diagnostic diagrams that are generally adopted to separate Low-Ionisation
Nuclear Emission-line Regions \citep[LINERs;][]{hec80} from Seyfert
galaxies \citep{kew06,but10,cid10,bal10}. However, for the higher redshift
samples, the observed wavelength range and the lower quality of the
spectra generally prohibit detection or measurement of some emission lines
required. Separation of the two populations has typically been performed
using either a single emission line flux ratio ($f_{\rm [OIII]
  5007}/f_{H\alpha}$ or $f_{\rm [OIII] 5007}/f_{\rm [OII] 3727}$) or a
single line equivalent width (EW$_{\rm [OIII]}$), with different authors
adopting slightly different criteria \citep[e.g.][]{lai94,jac97,tad98}.

\begin{figure*}
\begin{tabular}{cc}
\psfig{file=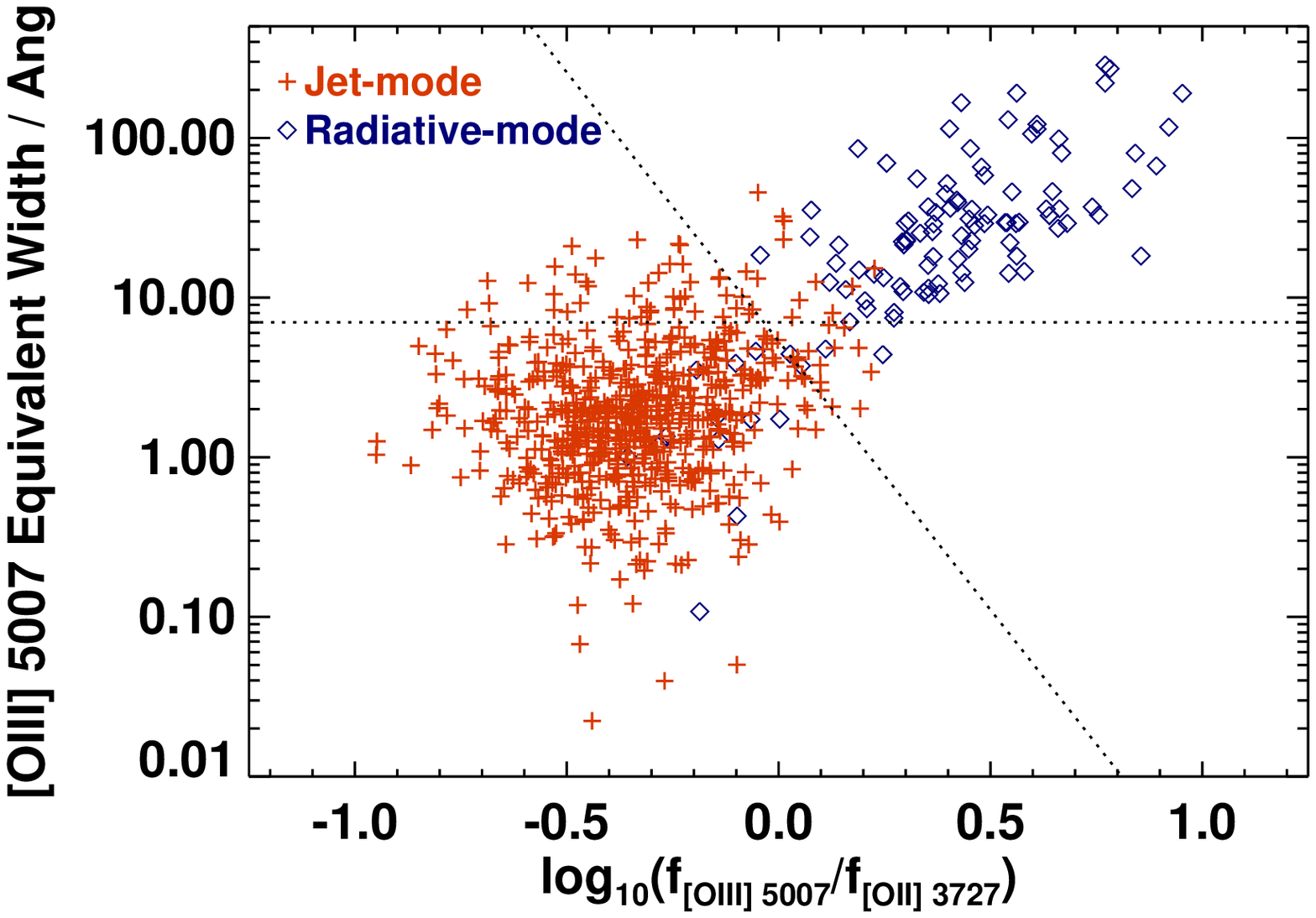,width=8cm,
clip=} 
&
\psfig{file=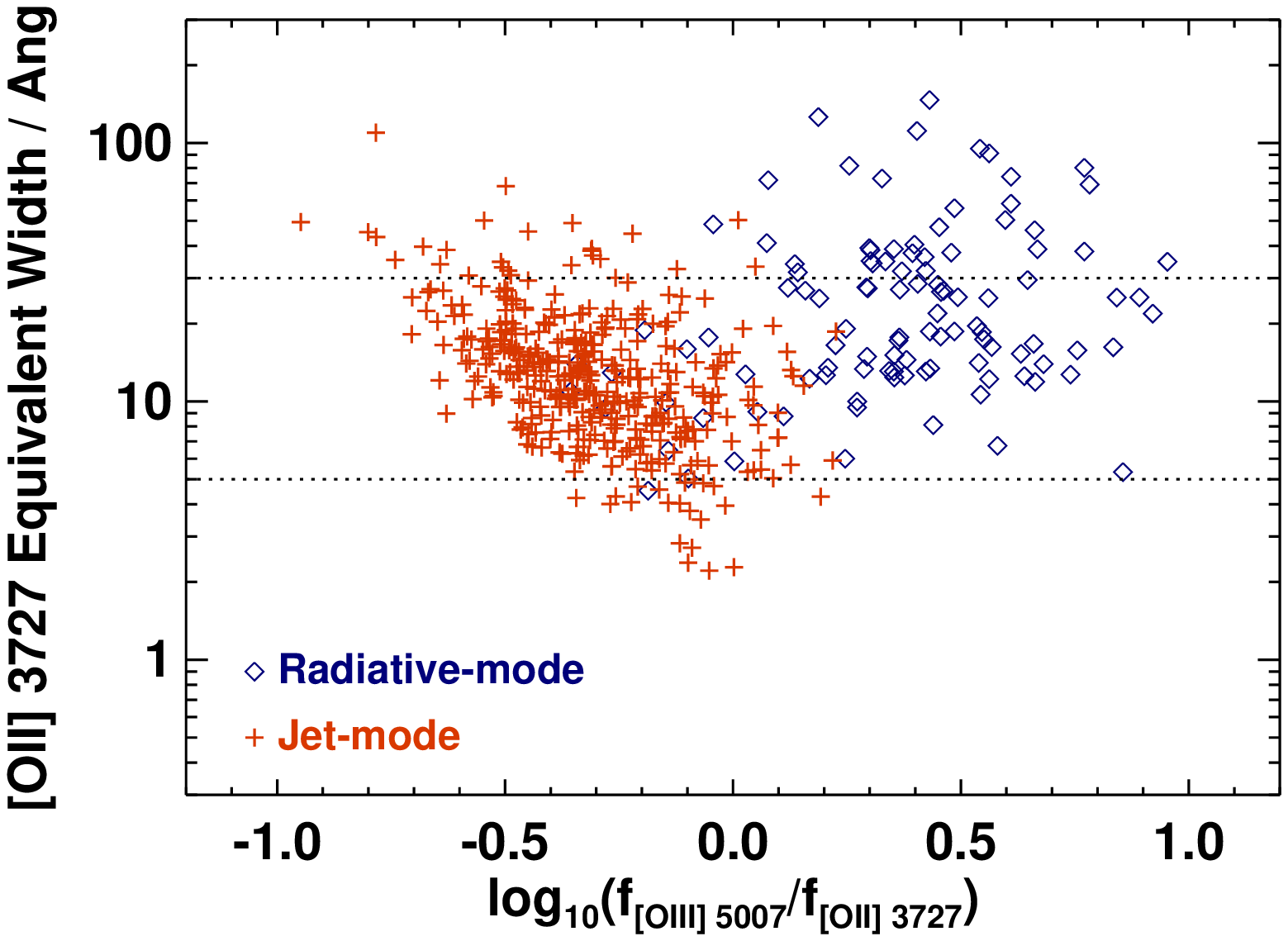,width=8cm,
clip=} \\
\end{tabular}
\caption{\label{fig_classify} The distribution of low-redshift
  radiative-mode and jet-mode AGN from BH12 on emission-line ratio versus
  equivalent width plots, for the [OIII]~5007 and [OII]~3727 emission
  lines. These are used to calibrate the classification division lines
  adopted for the high redshift samples (see Section~\ref{sec_classify}).}
\end{figure*}

For the $0.5 < z < 1.0$ sample, the [OII]~3727 line is available in all
optical spectra, and the [OIII]~5007 line in most. These lines therefore
form the basis of the classifications used in this paper.  The BH12 data
can be used to optimally calibrate the separation criteria using the line
flux ratio of these two lines, and their rest-frame equivalent widths.
For these analyses, the subsample of BH12 sources used was those with
[OII]~3727 and [OIII]~5007 emission lines detected with S/N $>$ 5 and
which were classified solely on the basis of emission line ratio
diagnostics. The left panel of Figure~\ref{fig_classify} shows a plot of
EW$_{\rm [OIII]}$ {\it vs} $f_{\rm [OIII] 5007} / f_{\rm [OII] 3727}$ for
this BH12 subsample, and demonstrates that where both emission lines are
available, the cleanest separation adopts a combination of these
parameters, rather than either individually. The division line adopted
here is 

\begin{displaymath}
{\rm log}_{10}\left({\rm EW}_{\rm [OIII]}\right) = 0.7 - 3.3 {\rm
  log}_{10}\left(\frac{f_{\rm [OII] 5007}}{f_{\rm [OIII] 3727}}\right).
\end{displaymath}

\noindent The parameters of this division line were derived by minimising
the quantity $f_{\rm wrong} = \left(f_{\rm jet-wrong}^2 + f_{\rm
  rad-wrong}^2\right)^{1/2}$ where $f_{\rm jet-wrong}$ and $f_{\rm
  rad-wrong}$ are the fraction of wrongly-classified jet-mode and
radiative-mode AGN respectively.

This classification criterion is adopted for spectra where flux
measurements of both lines are available. If only the [OIII] line is
available, or the spectra are not flux calibrated, then the classification
was made solely on EW$_{\rm [OIII]}$, with a division at 7\AA\ (again, this
value was derived by minimising $f_{\rm wrong}$). Finally, if only the [OII]
line is available, then the division was made using EW$_{\rm [OII]}$. As
shown in the right panel of Figure~\ref{fig_classify}, the two populations
overlap significantly in that parameter. Therefore objects with EW$_{\rm
  [OII]} > 30$\AA\ were classified as radiative-mode AGN, those with
EW$_{\rm [OII]} < 5$\AA\ were classified as jet-mode AGN, but those with
equivalent widths between these two limits were considered unclassifiable.

An extensive literature search was carried out to locate spectra for
sources within the samples used. For sources with available spectra in
electronic form, or with tabulated line properties, the above criteria
were applied. For some sources the only available spectra were in paper
form: most of these were powerful radio sources with very strong lines
whose classification as radiative-mode was unambiguous, in which case they
were classified by eye. The remainder were left unclassified. Some sources
which lacked either a spectrum or a classification were targeted in the
new WHT spectroscopic observations (see Appendix~\ref{app_specdetails})
and were classified on that basis. Note that for some sources spectra
exist but without a redshift having been determined. If these were of
sufficient quality to rule out the presence of an emission line with EW $>
5$\AA\ then the source was classified as a jet-mode AGN, otherwise it was
left unclassified. The final classifications for each source are listed in
Table~\ref{tab_sample}. In total it was possible to classify 123 sources
as radiative-mode and 46 sources as jet-mode, with 42 sources remaining
unclassifiable.

\section{The evolving radio luminosity functions of radio-AGN populations}
\label{sec_rlfs}

\subsection{Deriving the radio luminosity functions}
\label{sec_calcrlfs}

Radio luminosity functions were calculated using the standard technique,
$\rho = \sum_i 1/V_i$ \citep{sch68b,con89}, where $V_i$ is the volume
within which source $i$ could be detected. For the higher redshift
samples, the calculation of $V_i$ requires careful accounting of the
combination of different survey areas and depths, since sources detected
in one survey may have been detectable (and therefore have a contribution
to $V_i$) in another survey. For a given survey, a source of given
luminosity and spectral index is detectable out to the redshift ($z_{\rm
  lim}$) at which its radio flux density drops below the flux limit of
that survey. If the RLF is being calculated within a redshift range
$z_{\rm min} < z < z_{\rm max}$ (typically $0.5 < z < 1.0$ in this paper),
then: (i) if $z_{\rm lim} < z_{\rm min}$ the source could not be detected
by this survey in the redshift range studied; (ii) if $z_{\rm lim} >
z_{\rm max}$ then the source could be detected over the entire volume
probed by that survey between $z_{\rm min}$ and $z_{\rm max}$; (iii) if
$z_{\rm min} \le z_{\rm lim} \le z_{\rm max}$ then the source could be
detected over the subset of the volume between $z_{\rm min}$ and $z_{\rm
  lim}$. The total $V_i$ for each source is calculated by summing the
contributions to the detectable volume from all of the eight surveys,
taking account of any overlapping sky areas.

The RLFs were also parameterised with broken power-law fits ($\rho =
\rho_0 / [(L/L_0)^{\beta} + (L/L_0)^{\gamma}]$). These parameterised fits
were determined using a maximum-likelihood analysis \citep[cf.][]{mar83},
specifically by minimising the function

\begin{displaymath}
S = -2 \sum_{i=1}^N \ln \rho(L_i,z_i,\alpha_i) 
\end{displaymath}
\vspace*{-0.5cm}

\begin{displaymath}
\qquad\qquad + 2 \int\!\!\!\int\!\!\!\int
\rho(L,z,\alpha) \frac{{\rm d}V(L,z,\alpha)}{{\rm d}z}{\rm d}z\, {\rm
  d}(\log_{10}L)\, {\rm d}\alpha
\end{displaymath}

\noindent where $\rho(L,z,\alpha)$ is the space density of sources of
luminosity $L$ and spectral index $\alpha$ at redshift $z$, and
$\frac{{\rm d}V(L,z,\alpha)}{{\rm d}z}$ is the co-moving volume available
between redshift $z$ and $z + {\rm d}z$ for sources of luminosity $L$ and
spectral index $\alpha$, taking into account the sky areas and flux
density limits of the different constituent surveys. The first term is
therefore the sum of $\ln \rho$ over the $N$ sources in the sample, while
the second term integrates the model distribution and should evaluate to
approximately $2N$ for good fits.  

The distribution in spectral index was assumed to be independent of both
radio luminosity and redshift (over the narrow redshift range sampled),
and was evaluated as a Gaussian centred on $\alpha = 0.75$, with standard
deviation 0.15, cutting to zero below $\alpha = 0.5$ due to the
steep-spectrum selection limit. Tests indicate that the results are
unaffected if other sensible choices are adopted instead. For fitting of
the RLF broken power-law parameters at different redshifts, $\rho$ was
assumed to be independent of redshift within each studied redshift
bin. For some fits the value of $\gamma$ was fixed at 1.7 (consistent with
the best-fit values, and a reasonable fit in all cases) to ease the
degeneracies between the different parameters.

Marginalised errors on the parameter values were derived from the
covariance matrix, $\sigma^2_{\rm marg,i} = ([H]^{-1})_{ii}$ where $H_{ij}
= \frac{-\partial^2 S}{\partial p_i \partial p_j}$ is the Hessian matrix
(for parameters $p_i$) which was evaluated numerically. As well as these
marginalised errors, however, a significant source of uncertainty arises
from the presence of unclassified sources. To account for these, the
maximum likelihood analysis was carried out 1000 times, each time randomly
including or excluding each unclassified source (with equal
probability). The best-fit value for each parameter was determined by
taking the mean of these 1000 analyses. The uncertainty on the parameter
value was derived by combining the mean value of the marginalised error
for that parameter in quadrature with the standard deviation of the
parameter values determined from the 1000 iterations of the analysis. In
general the marginalised error was the dominant source of error,
indicating that small sample size and parameter degeneracies were more
important sources of error than the missing classifications.

\subsection{Radio luminosity functions results}
\label{sec_rlfres}

\begin{figure}
\begin{center}
\psfig{file=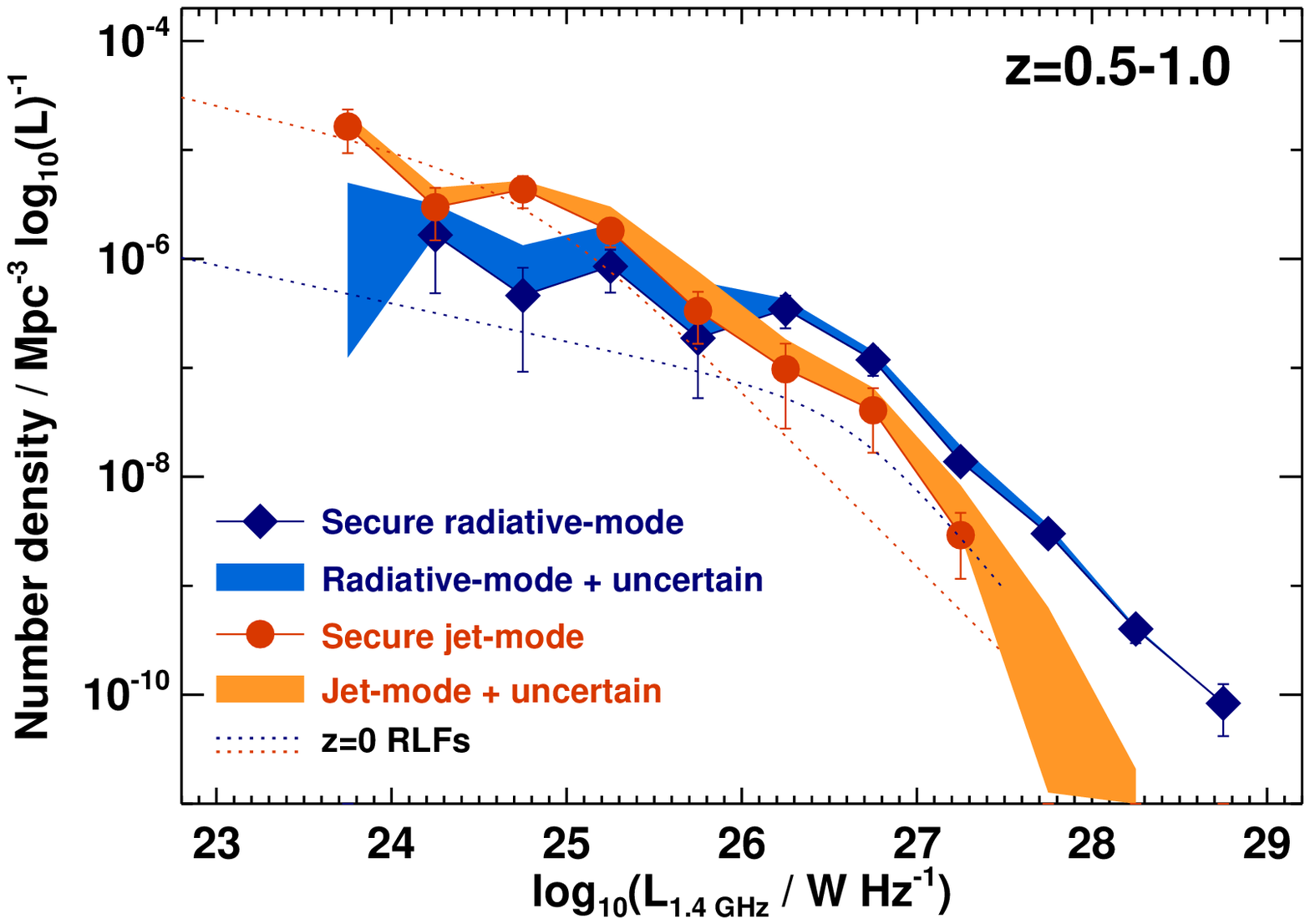,width=8.6cm,clip=}
\\
\psfig{file=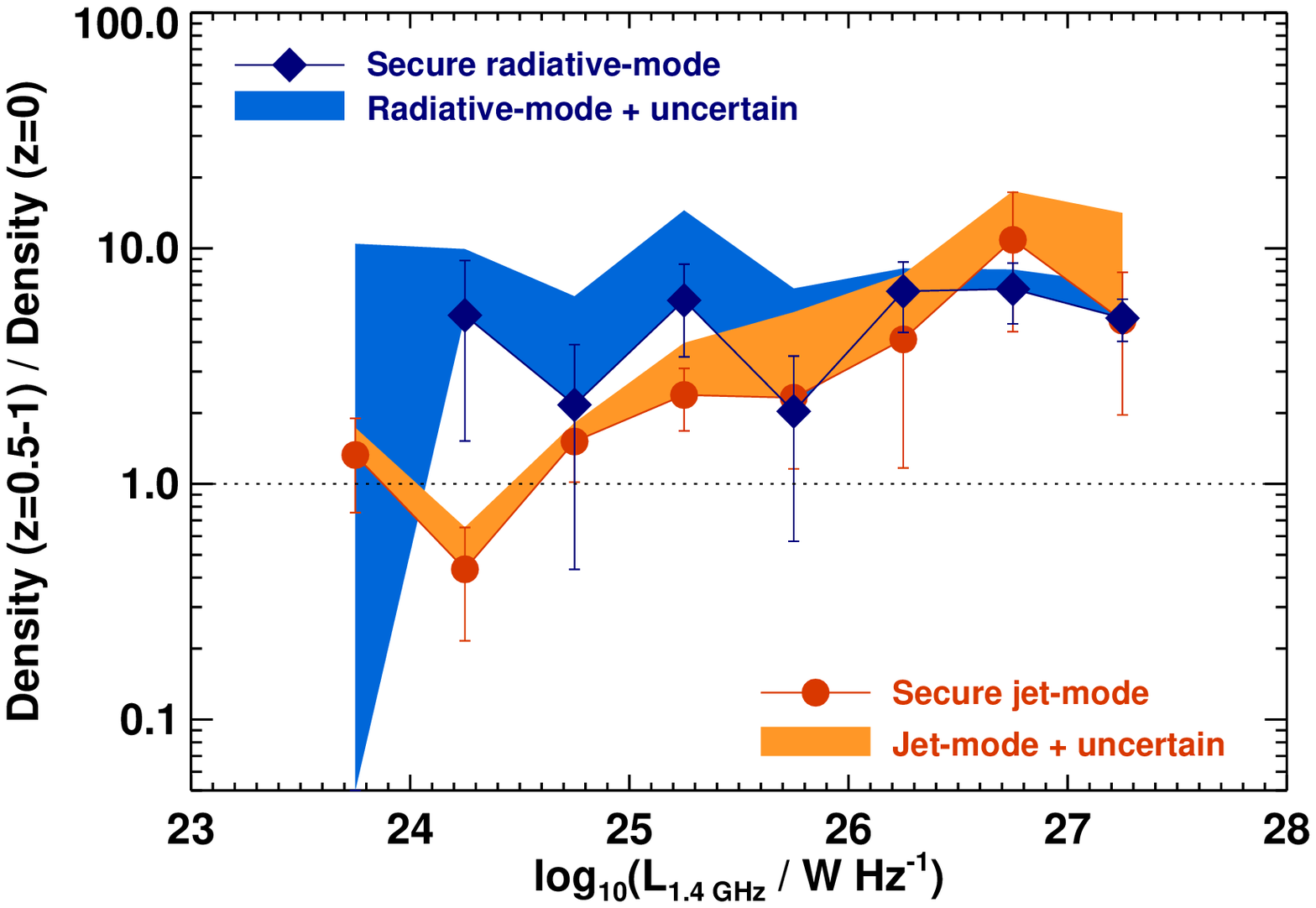,width=8.6cm,clip=}
\end{center}
\caption{\label{fig_rlfherglerg} Top: the RLFs of jet-mode and
  radiative-mode radio-AGN at $0.5 < z < 1.0$, compared with the local
  RLFs of the same populations. Bottom: the ratio of the space density of
  jet-mode and radiative-mode AGN at $0.5<z<1.0$, compared to the local
  Universe, as a function of radio luminosity. In both panels the data
  points and associated error bars represent the measured space density and
  its Poisson uncertainty, based solely on the securely-classified objects,
  while the shaded region represents the potential increase in space
  density arising from inclusion of all unclassified objects.}
\end{figure}

\begin{table*}
\caption{\label{tab_lumfunc} The $0.5 < z < 1.0$ radio luminosity
  functions at 1.4\,GHz, derived separately for the jet-mode and
  radiative-mode populations. Column 1 indicates the range of 1.4\,GHz
  radio luminosities of each bin. Column 2 gives the total number of radio
  sources in that radio luminosity range and Column 3 gives the associated
  space density, in units of number per log$_{10} L$ per Mpc$^3$. Columns
  4 to 9 show the equivalent values for radiative-mode radio-AGN, jet-mode
  radio-AGN and unclassified radio-AGN, respectively. Uncertainties are
  statistical Poissonian uncertainties only (see also
  Figure~\ref{fig_rlfherglerg}). Note that one jet-mode source has $L_{\rm 1.4
    GHz} < 10^{23.5}$W\,Hz$^{-1}$, so the number of sources in the table
  sums to only 210.}
\begin{tabular}{crcrcrcrc}
\hline
${\rm log} L_{\rm 1.4 GHz}$&
\multicolumn{2}{c}{All radio sources} &
\multicolumn{2}{c}{Radiative-mode} &
\multicolumn{2}{c}{Jet-mode} &
\multicolumn{2}{c}{Unclassified} 
\\
W Hz$^{-1}$  &
N~ & ${\rm log}_{10} \rho$ &
N~ & ${\rm log}_{10} \rho$ &
N~ & ${\rm log}_{10} \rho$ &
N~ & ${\rm log}_{10} \rho$ \\
\hline
23.5-24.0 & 10 &  -$4.67^{+0.13}_{-0.19}$ &  0 &                        &  6 & -$4.79^{+0.16}_{-0.24}$ &  4 &    -$5.30$ \\
24.0-24.5 &  8 &  -$5.21^{+0.13}_{-0.19}$ &  2 & -$5.78^{+0.23}_{-0.53}$ &   4 & -$5.53^{+0.18}_{-0.30}$ & 2 &    -$5.82$ \\
24.5-25.0 & 13 &  -$5.25^{+0.11}_{-0.15}$ &  1 & -$6.34^{+0.26}_{-0.70}$ &  10 & -$5.36^{+0.12}_{-0.17}$ & 2 &    -$6.06$ \\
25.0-25.5 & 25 &  -$5.41^{+0.08}_{-0.10}$ &  6 & -$6.07^{+0.15}_{-0.24}$ &  12 & -$5.74^{+0.11}_{-0.15}$ & 7 &    -$5.92$ \\
25.5-26.0 & 11 &  -$6.02^{+0.12}_{-0.16}$ &  2 & -$6.73^{+0.24}_{-0.55}$ &   4 & -$6.48^{+0.18}_{-0.30}$ & 5 &    -$6.36$ \\
26.0-26.5 & 14 &  -$6.28^{+0.11}_{-0.14}$ & 10 & -$6.46^{+0.12}_{-0.18}$ &   2 & -$7.01^{+0.23}_{-0.55}$ & 2 &    -$7.06$ \\
26.5-27.0 & 20 &  -$6.73^{+0.09}_{-0.12}$ & 14 & -$6.93^{+0.11}_{-0.15}$ &   3 & -$7.39^{+0.20}_{-0.39}$ & 3 &    -$7.60$ \\
27.0-27.5 & 45 &  -$7.65^{+0.07}_{-0.08}$ & 30 & -$7.86^{+0.08}_{-0.10}$ &   4 & -$8.54^{+0.20}_{-0.40}$ &11 &    -$8.27$ \\ 
27.5-28.0 & 43 &  -$8.44^{+0.07}_{-0.08}$ & 38 & -$8.52^{+0.07}_{-0.08}$ &   0 &                       & 5 &    -$9.20$  \\
28.0-28.5 & 17 &  -$9.37^{+0.10}_{-0.13}$ & 16 & -$9.40^{+0.10}_{-0.13}$ &   0 &                       & 1 &   -$10.68$ \\
28.5-29.0 &  4 & -$10.08^{+0.18}_{-0.30}$ &  4 & -$10.08^{+0.18}_{-0.30}$ &  0 &                       &  0 &           \\
\hline  
\end{tabular}
\end{table*}

\begin{table*}
\caption{\label{tab_lumfuncparams} Parameters of broken power law fits to
  the 1.4\,GHz radio luminosity functions, of the form $\rho = \rho_0 /
  [(L/L_0)^{\beta} + (L/L_0)^{\gamma}]$, where $\rho$ and
  $\rho_0$ are measured in units of number per log$_{10} L$ per
  Mpc$^3$. For some fits, the value of $\gamma$ is fixed at 1.7.}
\begin{tabular}{cccccc}
\hline
AGN-type & Redshift    & $L_0$ & log$_{10}(\rho_0)$ & $\beta$ & $\gamma$ \\ 
\hline
All      & $z < 0.3$   & $24.95 \pm 0.14$ & -$5.33 \pm 0.12$ & $0.42 \pm 0.04$ & $1.66 \pm 0.21$ \\
         & $0.5<z<1.0$ & $26.22 \pm 0.14$ & -$5.96 \pm 0.16$ & $0.45 \pm 0.06$ & $1.68 \pm 0.06$ \\
Jet-mode & $z < 0.3$   & $24.81 \pm 0.18$ & -$5.30 \pm 0.17$ & $0.39 \pm 0.06$ & $1.61 \pm 0.19$  \\
         & $0.5<z<1.0$ & $25.50 \pm 0.14$ & -$5.62 \pm 0.16$ & $0.35 \pm 0.12$ & 1.70 (fixed) \\
         & $0.5<z<0.7$ & $25.11 \pm 0.15$ & -$5.15 \pm 0.19$ & $0.28 \pm 0.14$ & 1.70 (fixed) \\
         & $0.7<z<1.0$ & $25.77 \pm 0.19$ & -$5.86 \pm 0.26$ & $0.13 \pm 0.24$ & 1.70 (fixed) \\
Radiative-mode&$z<0.3$ & $26.62 \pm 0.11$ & -$7.32 \pm 0.08$ & $0.35 \pm 0.02$ & 1.70 (fixed) \\
         & $0.5<z<1.0$ & $26.45 \pm 0.11$ & -$6.37 \pm 0.17$ & $0.30 \pm 0.10$ & 1.70 (fixed) \\
\hline
\end{tabular}
\end{table*}

The derived $0.5 < z < 1.0$ RLFs of the radiative-mode, jet-mode and
unclassified radio-AGN are tabulated separately in
Table~\ref{tab_lumfunc}.  The RLFs are also shown in the upper panel of
Figure~\ref{fig_rlfherglerg}: the data points indicate the RLFs of the
securely classified objects, with associated error bars, while the shaded
regions indicate the extent to which these might be increased by inclusion
of the unclassified objects. The parameters of the broken power-law fits
(calculated using the method of Section~\ref{sec_calcrlfs}) are given in
Table~\ref{tab_lumfuncparams}. The fits to the local RLFs are also shown
in Figure~\ref{fig_rlfherglerg}, from which the cosmic evolution of the
RLFs of each class can be seen. This is more clearly demonstrated in the
lower panel of Figure~\ref{fig_rlfherglerg} which shows the ratio of the
high-to-low redshift RLFs in terms of a space-density scaling factor as a
function of radio luminosity.

The radiative-mode radio-AGN evolve by a constant factor of $\approx 7$ in
co-moving space density, between the local Universe and $z \approx 0.75$,
at all radio luminosities.  At high radio luminosities (where these
sources dominate) this is entirely consistent with previous determinations
of the evolution of the total RLF \citep{dun90,rig11}. The RLF fits data
prefer a pure density evolution model, with little change in $L_0$,
although sufficient parameter degeneracy remains for the radiative-mode
AGN fitting that a combination of density and luminosity evolution cannot
be ruled out.

The evolution of the jet-mode radio-AGN is rather more complicated. At low
radio luminosities ($L_{\rm 1.4 GHz} \lta 10^{25}$W\,Hz$^{-1}$), these
show little or no cosmic evolution. This is in line with the previous
measurements of the low evolution of the RLF as a whole at these low
luminosities \citep{sad07,don09}, since the jet-mode AGN dominate the
overall population. Indeed, the mild evolution seen here in the total RLF
is mostly driven by the strong evolution of the sub-dominant
radiative-mode population. At higher radio luminosity, however, the
jet-mode AGN do show significant cosmic evolution, approaching that of the
radiative-mode AGN.

\begin{figure}
\centerline{
\psfig{file=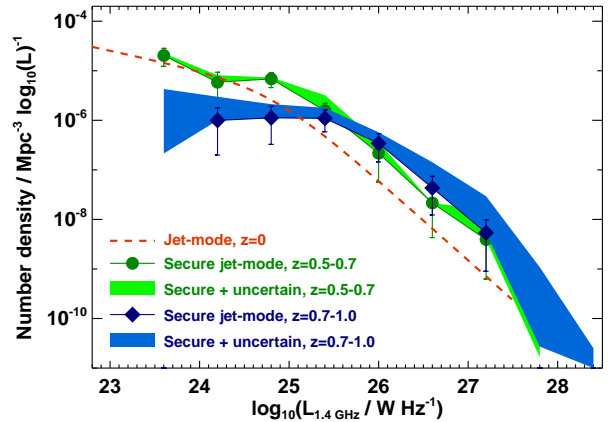,width=8.6cm,clip=}}
\caption{\label{fig_lergsplit} Top: the RLF of jet-mode radio-AGN at $0.5
  < z < 0.7$ and $0.7 < z < 1.0$ compared with that in the local
  Universe. Errors bars and shaded regions are as defined in
  Fig~\ref{fig_rlfherglerg}.}
\end{figure}

Figure~\ref{fig_lergsplit} considers the RLF of the jet-mode AGN, now
split into two redshift ranges: $0.5 < z < 0.7$ and $0.7 < z < 1.0$ (see
also Table~\ref{tab_lumfuncsplit}). It is evident that at the lowest radio
luminosities ($L_{\rm 1.4 GHz} \lta 10^{24}$W\,Hz$^{-1}$) the space
density of jet-mode AGN remains broadly constant out to $z \approx 0.5$
and then decreases\footnote{Note that, as is evident from
  Figure~\ref{fig_pzdist}, the fraction of unclassified objects is quite
  high at low luminosities in the higher redshift bin, in part because
  classification at these redshifts is based on [OII] alone. It seems
  likely that many of these unclassified sources will be jet-mode sources,
  and that the true space density of jet-mode sources will lie close to
  the upper envelope of the shaded region in
  Figure~\ref{fig_lergsplit}. The space density decline is therefore less
  pronounced than may first meet the eye.} to $z=1$. At moderate
luminosities ($10^{24}$W\,Hz$^{-1} \lta L_{\rm 1.4 GHz} \lta
10^{26}$W\,Hz$^{-1}$) the space density increases to $z \sim 0.5$ before
falling. At the highest luminosities, the space density continues to
increase with increasing redshift out to $z \sim 1$. This is consistent
with the luminosity-dependent evolution of the overall RLF seen by
\citet{rig11} but does indicate that the picture is more complicated than
just differential evolution of two different AGN populations.

\begin{table*}
\caption{\label{tab_lumfuncsplit} The 1.4\,GHz radio luminosity functions of
  jet-mode AGN split into $0.5<z<0.7$ and $0.7<z<1.0$ redshift
  bins. Column 1 indicates the range of 1.4\,GHz radio luminosities of
  each bin. Columns 2-5 show the results for the $0.5<z<0.7$ bin, with the
  columns indicating respectively the number of jet-mode AGN, their space
  density (in units of number per log$_{10} L$ per Mpc$^3$), the number of
  unclassified sources, and their space density. Columns 6-9 repeat these
  results for the higher redshift range.}
\begin{tabular}{crcrcrcrc}
\hline
${\rm log} L_{\rm 1.4 GHz}$&
\multicolumn{4}{c}{..............$0.5 < z < 0.7$..............} &
\multicolumn{4}{c}{..............$0.7 < z < 1.0$..............} \\
W Hz$^{-1}$ & 
\multicolumn{2}{c}{Jet-mode} &
\multicolumn{2}{c}{Unclassified} &
\multicolumn{2}{c}{Jet-mode} &
\multicolumn{2}{c}{Unclassified} 
\\
&
N~ & ${\rm log}_{10} \rho$ &
N~ & ${\rm log}_{10} \rho$ &
N~ & ${\rm log}_{10} \rho$ &
N~ & ${\rm log}_{10} \rho$ \\
\hline
23.3-23.9 &  7 & -$4.69^{+0.14}_{-0.22}$ &    1 & -$5.65$  &   0 &                         &   2  &   -$5.37$ \\
23.9-24.5 &  3 & -$5.23^{+0.20}_{-0.39}$ &    1 & -$5.65$  &   1 & -$6.00^{+0.26}_{-0.70}$  &   2  &   -$5.70$ \\ 
24.5-25.1 & 10 & -$5.16^{+0.13}_{-0.18}$ &    1 & -$6.33$  &   2 & -$5.95^{+0.23}_{-0.54}$  &   1  &   -$6.00$ \\
25.1-25.7 &  6 & -$5.81^{+0.15}_{-0.23}$ &    6 & -$5.78$  &   5 & -$5.96^{+0.17}_{-0.27}$  &   4  &   -$6.16$ \\ 
25.7-26.3 &  2 & -$6.66^{+0.24}_{-0.58}$ &    1 & -$6.85$  &   3 & -$6.47^{+0.20}_{-0.37}$  &   2  &   -$6.64$ \\
26.3-26.9 &  1 & -$7.67^{+0.26}_{-0.70}$ &    0 &          &   2 & -$7.36^{+0.23}_{-0.55}$  &   3  &   -$7.00$ \\
26.9-27.5 &  2 & -$8.41^{+0.26}_{-0.80}$ &    3 & -$7.93$  &   2 & -$8.27^{+0.26}_{-0.78}$  &   9  &   -$7.62$ \\
\hline                                                                                 
\end{tabular}
\end{table*}

\subsection{Modelling the jet-mode RLF evolution}
\label{sec_jetmodels}

\subsubsection{A pure density evolution model}
\label{sec_moda}

In the simplest picture of the jet-mode radio-AGN population, these AGN
are hosted by quiescent galaxies living within hot gas haloes, in which
star-formation has been largely extinguished, and the AGN is fuelled by
the cooling of the hot gas (see Section~\ref{sec_intro}). In this picture,
it is possible to predict the evolution in the space density of jet-mode
AGN from the evolution of potential host galaxies. In recent years there
have been a number of observational determinations of the stellar mass
function of quiescent galaxies both in the local Universe
\citep[e.g.][]{bal12} and out to high redshifts
\citep[e.g.][]{dom11,mou13,ilb13,muz13}. These stellar mass functions can
be combined with the prevalence of jet-mode AGN activity as a function of
stellar mass \citep[$f_{\rm AGN} \approx 0.01 (M_* /
  10^{11}M_{\odot})^{2.5}$; ][]{bes05b,jan12} to predict the evolution of
the space density of jet-mode AGN as a function of redshift.

\begin{figure}
\centerline{
\psfig{file=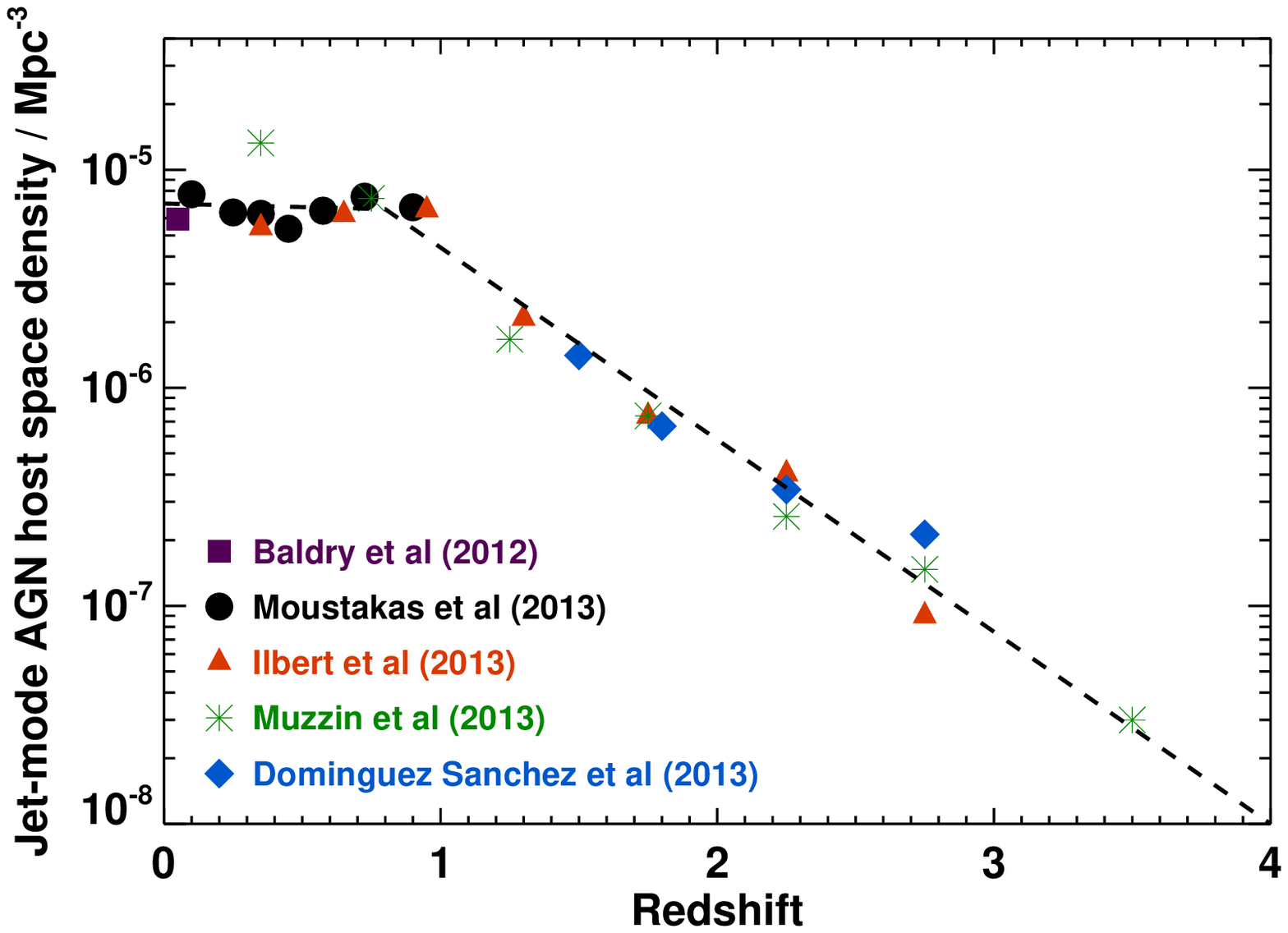,width=8.3cm,clip=}}
\caption{\label{fig_hostevol} Cosmic evolution of the space density of
  potential hosts of jet-mode AGN, derived by combining the stellar mass
  functions of quiescent galaxies at different redshifts
  \citep[from][]{dom11,bal12,mou13,ilb13,muz13} with the prevalence of
  jet-mode AGN activity as a function of stellar mass \citep[$f_{\rm AGN}
    \approx 0.01 (M / 10^{11}M_{\odot})^{2.5}$; ][]{bes05b,jan12}. All
  quiescent galaxy stellar mass functions are first converted onto a
  Chabrier IMF. The results from \citet{mou13} are scaled down by a factor
  of two, and those of \citet{dom11} by a factor of 1.5 to bring the
  different datasets into agreement for visualisation -- but the form of
  the fitted redshift evolution (dashed line) is consistent across all
  datasets.}
\end{figure}

Figure~\ref{fig_hostevol} shows the result of this analysis. To derive
this, the literature mass functions were shifted in mass (where necessary)
to move then all onto a Chabrier IMF. Furthermore, to bring different
datasets in to line with each other for visualisation purposes
(differences are likely to be due to different definitions of quiescent
galaxies), it was necessary to vertically shift the data points of
\citet{mou13} down by a factor of two, and those of \citet{dom11} down by
a factor of 1.5. These corrections means that the absolute values of the
plotted space densities may be unreliable, but the trends with redshift
are robust, as these are consistent across all datasets. The space density
of jet-mode AGN hosts is modelled in a simple manner as evolving as
$(1+z)^{-0.1}$ out to redshift $z=0.8$, and then as $(1+z)^{-6.5}$ at
higher redshifts. A similar result is obtained if one instead simply
considers the evolution of the space density of all quiescent galaxies
more massive than $10^{10} M_{\odot}$.

\begin{figure}
\centerline{
\psfig{file=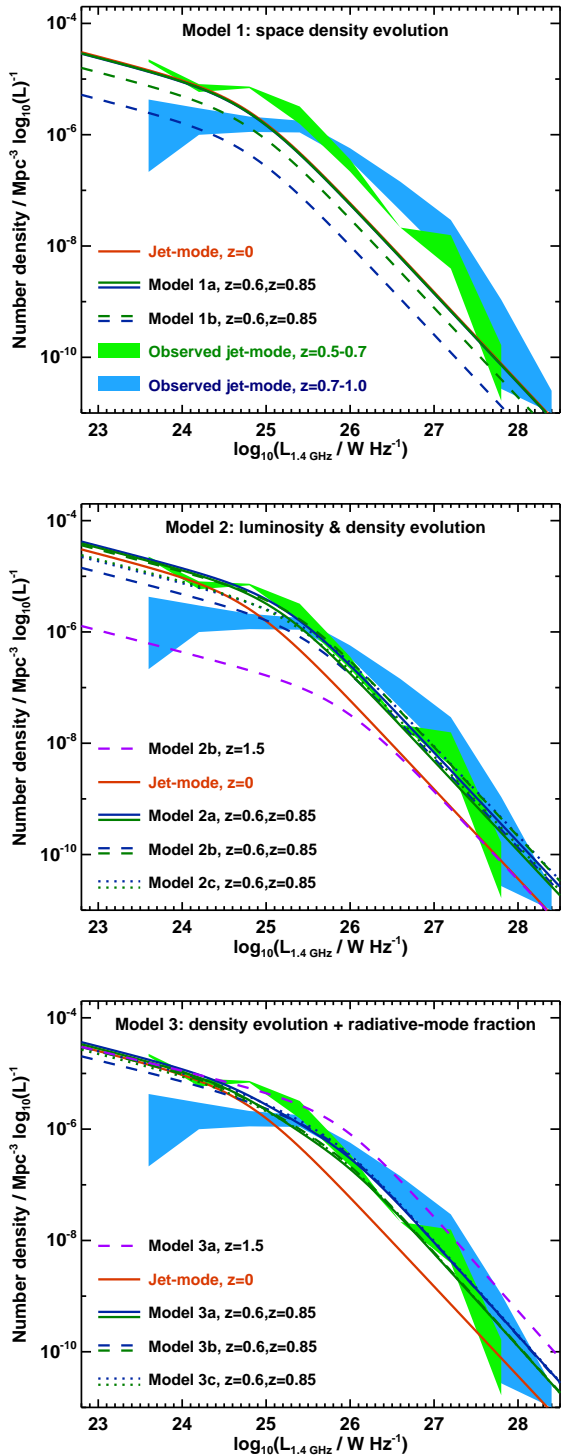,width=8.0cm,clip=}}
\caption{\label{fig_jetmodemodels} A comparison of the observed jet-mode
  RLFs with model predictions. In all panels the green and blue shaded
  regions represent the uncertainty on the jet-mode RLFs at $0.5<z<0.7$
  and $0.7<z<1.0$ respectively, as shown in Fig.~\ref{fig_lergsplit},
  while the various green and blue lines show the model predictions. The
  top panel shows the results for versions of Model 1 (space density
  evolution; see \S\ref{sec_moda} and \S\ref{sec_modb}), the middle panel
  shows versions of Model 2 (luminosity density evolution;
  \S\ref{sec_mod2}) and the lower panel shows versions of Model 3 (density
  evolution with radiative-mode contribution; \S\ref{sec_mod3}). In all
  panels the red solid line shows the $z=0$ jet-mode RLF for
  comparison. The purple dashed lines show extrapolations of Models 2b and
  3a to $z=1.5$. Note that in the top panel the solid lines are largely on
  top of each other. }
\end{figure}

Under this simplest picture of jet-mode AGN evolution, the jet-mode RLF
will demonstrate pure density evolution, evolving down in space density in
accordance with the cosmic evolution of the potential host galaxies, just
derived. A comparison between the data and this simplest model prediction
is shown as Model 1a in Figure~\ref{fig_jetmodemodels}.

\subsubsection{Delays in the onset of jet-mode AGN activity}
\label{sec_modb}

A remarkable discovery over the last decade is that massive galaxies at
high redshifts are significantly more compact than those of the same mass
in the nearby Universe \citep[e.g.][]{dad05}. However, the host galaxies
of powerful radio sources at moderate to high redshifts are as large as
those nearby \citep[e.g.][]{bes98d}. Lower power radio-AGN are also found
to be hosted by galaxies that are larger than other ellipticals of the
same mass (Caldwell et~al. in prep.). A plausible explanation is that
after massive galaxies have their star-formation quenched there is a
time-delay before the surrounding hot halo has established itself into a
state where gas cooling and AGN fuelling can proceed, and that this time
delay is long enough for the processes that `puff up' the galaxy to have
occurred. For example, if the transition to a quenched state is driven by
a powerful quasar outburst removing most of the cold gas from the galaxy,
then that same outburst might disturb the surrounding hot gas and lead to
a delay before gas cooling established. The cooling time of gas at
$\approx 10$\,kpc radii in massive elliptical galaxies is typically of
order a Gyr \citep[e.g.][]{pan14}, which interestingly is broadly similar
to the timescale for the `puffing up' of a galaxy in the quasar feedback
model of \citet{fan08}, where the removal of gas is argued to induce an
expansion of the stellar distribution over a few tens of dynamical times
($\approx 2$\,Gyr). Other mechanisms for increasing the sizes of galaxies
(e.g.\ multiple minor mergers) may work on similar timescales.

In order to account for this possibility, Model 1a was adapted to include
a time delay $\tau = 2$\,Gyr between the formation of quiescent galaxies
and their ability to produce jet-mode radio-AGN. (Note that in later
models, $\tau$ is allowed to be a free parameter; however in this first
simple model, the fit to the data is sufficiently poor that meaningless
results are obtained, so the 2\,Gyr value is used solely for illustrative
purposes). In practice this time delay was incorporated by considering the
space density of jet-mode radio-AGN at redshift $z$ to evolve as the space
density of potential jet-mode host galaxies (from
Figure~\ref{fig_hostevol}) at redshift $z'$, where redshift $z'$ is the
redshift at which the Universe was $\tau$ younger than at redshift
$z$. This prediction is shown as Model 1b in
Figure~\ref{fig_jetmodemodels}.

\subsubsection{Luminosity-density evolution of jet-mode AGN}
\label{sec_mod2}

These pure density evolution models are clearly unable to explain the
behaviour at high radio luminosities, where the space density of jet-mode
AGN increases with increasing redshift. This problem may be resolved if
the radio luminosities of the sources systematically increase with
redshift. Physically, this can be understood as follows. For a given
jet-power, the synchrotron luminosity of a source depends strongly on the
density of the environment into which it is expanding: in higher density
environments the radio lobes remain more confined and adiabatic expansion
losses are lower, leading to higher synchrotron luminosities
\citep[e.g.][]{bar96a}. At higher redshifts the average density of the
Universe is higher, and also the gas fraction is higher. Each of these
could plausibly lead to an increase in the radio luminosity with
redshift. In Model 2, therefore, the characteristic luminosity ($L_0$) of
the RLF is allowed to evolve as $(1+z)^\delta$, with $\delta$ a free
parameter. Model variations 2a, 2b and 2c are considered.  In Model 2a,
this luminosity evolution is combined with the space density evolution of
potential host galaxies at that redshift, as in Model 1a. In Model 2b, the
luminosity evolution is combined with the space density evolution of
potential hosts, including the time-lag of Model 1b, but allowing the time
lag $\tau$ to be a free parameter. In Model 2c, the luminosity evolution
is combined with a simple parameterised space density evolution of $\rho_0
\propto (1+z)^\eta$, with $\eta$ a free parameter.

To derive the best-fitting values of the parameters of these models, the
maximum likelihood analysis described in Section~\ref{sec_calcrlfs} was
used, again with 1000 Monte-Carlo iterations for the inclusion or
exclusion of unclassified sources. The best-fitting values of each
parameter, and their uncertainties, are shown in
Table~\ref{tab_evoljet}. The resultant model predictions for the evolution
of the jet-mode RLFs are shown in Figure~\ref{fig_jetmodemodels}.

\begin{table*}
\caption{\label{tab_evoljet} Best-fit parameter values, and their
  uncertainties, for the modelling of the jet-mode radio-AGN RLF.  The
  space density of the RLF at redshift $z$ is modelled as declining either
  with the available space density of potential hosts at redshift $z$
  (version `a'; \S\ref{sec_moda}), or as the space density of potential
  hosts at an earlier redshift $z'$, where $z'$ is the redshift
  corresponding to a time $\tau$ before redshift $z$ (version `b';
  \S\ref{sec_modb}), or declining as $(1+z)^\delta$ (version `c'). In
  versions of Model 2, the characteristic luminosity of the RLF also
  evolves as $(1+z)^\eta$ (\S\ref{sec_mod2}). In versions of Model 3 there
  is instead an additional contribution from radiative-mode AGN hosts,
  modelled as the radiative-mode RLF at that redshift scaled up by factors
  $f_{\rho}$ in space density and $f_{\rm L}$ in luminosity
  (\S\ref{sec_mod3}).}
\begin{tabular}{ccccccc}
\hline
Model    & Space density & $\delta$ & $\tau$ & $\eta$ & $f_{\rho}$ & $f_{\rm L}$ \\ 
         &  evolution    &          & [Gyr]  & & & \\
\hline
Model 1a & As potential hosts             &    ---         &    ---        & ---           &      ---      &    ---          \\
Model 1b & As potential hosts, with delay &    ---         & 2.0 (fixed)   & ---           &      ---      &    ---          \\
Model 2a & As potential hosts             &    ---         &    ---        & $1.6 \pm 0.2$ &      ---      &    ---          \\
Model 2b & As potential hosts, with delay &    ---         & $1.5 \pm 0.2$ & $2.8 \pm 0.2$ &      ---      &    ---          \\
Model 2c & $\rho_0 \propto (1+z)^\delta$   & $-1.6 \pm 0.3$ &    ---        & $2.8 \pm 0.3$ &      ---      &    ---          \\
Model 3a & As potential hosts             &    ---         &    ---        &      ---      & $1.2 \pm 0.4$ & $0.18 \pm 0.04$ \\
Model 3b & As potential hosts, with delay &    ---         & $1.4 \pm 0.3$ &      ---      & $2.0 \pm 0.7$ & $0.14 \pm 0.03$ \\
Model 3c & $\rho_0 \propto (1+z)^\delta$   & $-0.9 \pm 0.5$ &    ---        &      ---      & $2.0 \pm 0.8$ & $0.14 \pm 0.04$ \\
\hline
\end{tabular}
\end{table*}

\subsubsection{A radiative-mode contribution to the jet-mode AGN}
\label{sec_mod3}

An alternative explanation can also be considered for the short-comings of
Model 1 at high radio luminosities. At high radio luminosities the overall
RLF is dominated by the radiative-mode sources. It is possible that a
subset of the sources classified as jet-mode sources are not truly
quiescent galaxies fuelled by cooling of gas within hot gas haloes, but
rather are related to the radiative-mode radio-AGN population. At the
simplest level this could be due to mis-classification of some sources,
although this would require significant cosmic evolution in the EW$_{\rm
  [OIII]}$ division line between the populations. More plausible, this
could be to do with the physical properties of a `jet-mode' source.

BH12 argued that the distinction between radiative-mode and jet-mode
activity was primarily down to the Eddington-scaled accretion rate on to
the black hole: for Eddington-scaled accretion rates above about 1\%, a
geometrically-thin, luminous accretion disk forms and the AGN is
classified as radiative-mode; at lower Eddington-scaled accretion rates
there is instead a geometrically-thick radiatively-inefficient accretion
flow in which the energetic output of the AGN is primarily in the form of
powerful radio jets -- a jet-mode source. AGN powered by the cooling of
gas from hot haloes will invariably be fuelled at relatively low accretion
rates and thus will be jet-mode sources. Sources fuelled by cold dense gas
are capable of much higher Eddington-scaled accretion rates and therefore
can appear as radiative-mode AGN. However, accretion onto the AGN is a
stochastic process and it would be unsurprising if at some times cold-gas
fuelling occurred at rates below the critical Eddington fraction, leading
to a changed accretion mode and a jet-mode classification.

In the nearby Universe the jet-mode radio-AGN population is dominated by
the hot-gas fuelled sources (cf. BH12). Thus, the ``jet-mode versus
radiative-mode'' and ``hot-gas-fuelled versus cold-gas-fuelled''
distinctions are largely synonymous. Towards higher redshifts, however,
the prevalence of hot-gas fuelled sources will fall (due to fewer
potential hosts) and that of cold-gas fuelled sources rises (due to higher
gas availability) and so cold-gas-fuelled sources may begin to make a
significant contribution to the jet-mode population. In this respect it is
interesting that \citet{jan12} found that high power jet-mode AGN are more
likely to be blue in colour, i.e. star forming, which would fit this
picture.

To characterise this in a simple manner, the high-redshift jet-mode RLF
can be modelled as being composed of two populations. The first population
is the genuine hot-gas-fuelled sources and is constructed by evolving the
local jet-mode RLF with pure density evolution due to the decreasing space
density of potential host galaxies (evolving with variants a, b and c, as
for Model 2 above). To this is added a radiative-mode contribution, which
is modelled as being the radiative-mode RLF at the relevant redshift,
scaled in space density by a factor $f_{\rho}$ and in luminosity by a
factor $f_{\rm L}$. The luminosity scaling accounts for the accretion
rates being lower at the times that these galaxies are classified as
jet-mode. The density scaling factor accounts both for the fact that not
all radiative-mode AGN host galaxies may go through jet-mode phases, and
for the relative durations of radiative-mode and jet-mode radio
phases. Note that $f_{\rho}$ is allowed to be greater than unity, if the
jet-mode phase is longer lived.

Once again, Monte-Carlo iterations of the maximum likelihood analysis were
employed to derive the best-fit parameters for these models, Models 3a, 3b
and 3c. The best-fitting values of each parameter are shown in
Table~\ref{tab_evoljet} and the model predictions are shown in
Figure~\ref{fig_jetmodemodels}.

\section{Discussion}
\label{sec_discuss}

\subsection{Radiative-mode radio-AGN}

The evolution in the space density of radiative-mode radio-AGN (a factor
$\sim 7$ from the local Universe to $z \sim 0.75$) is remarkably similar
to the amount by which the cosmic star formation rate density has
increased over the same cosmic interval \citep[e.g.][and references
  therein]{sob13,mad14}. It is also comparable to the evolution of the
quasar luminosity function \citep[ie.\ radio-quiet radiative-mode AGN;
][]{has05,hop07,cro09}, although a combination of density and luminosity
evolution is usually preferred for the latter. These results are
consistent with the picture whereby both the radiative-mode AGN and
star-formation activity are simply controlled by the availability of a
supply of cold gas to the galaxy \citep[e.g.\ see discussion in][]{hec14}.

\subsection{The jet-mode RLF}

At low radio luminosities ($L_{\rm 1.4 GHz} < 10^{25}$W\,Hz$^{-1}$) the
jet-mode RLF shows a gradual decline with increasing redshift, which can
be explained by a decrease in the space density of available host galaxies
(Model 1). Thus, previous analyses of the prevalence of radio-AGN as a
function of stellar mass out to $z \sim 1$ \citep{tas08,don09,sim13},
which have been dominated by sources at these luminosities, have found
results broadly in line with those of the local Universe.

At high radio luminosities, however, an increase in space density with
increasing redshift is seen. The results of Figure~\ref{fig_jetmodemodels}
indicate that variants of both Models 2 and 3 are able to account for
this, although Model 2 does so with one fewer free parameter. Amongst the
Model 2 options, Model 2b provides the best match to the observed
data. This luminosity-density evolution model requires the jet-mode AGN
luminosities to scale as $(1+z)^{2.8}$ and adopts a 1.5\,Gyr delay between
the creation of a quiescent galaxy and the onset of jet-mode radio-AGN
activity due to the cooling of hot gas from the halo. It is interesting
that this time delay is in line with the typical cooling time of gas at
$\approx 10$\,kpc radii in massive elliptical galaxies
\citep[e.g.][]{pan14} which might provide the AGN fuel source, and also
with the $\approx 2$\,Gyr dynamical expansion timescale calculated in the
quasar-feedback model of \citet{fan08}.

There is little to distinguish between the various Model 3 options. Each
predicts $f_{\rm L} \approx$ 0.1-0.2 implying that, if this model is
correct, cold-gas-fuelled AGN scale down in luminosity by nearly an order
of magnitude (due to a corresponding decrease in accretion rate) as they
transition from radiative-mode to jet-mode AGN activity. This value makes
sense physically, as it is the decrease required to take a radiative-mode
AGN (typical $L/L_{\rm Edd} \sim 0.1$) down into the advection-dominated
accretion flow regime ($L/L_{\rm Edd} \lta 0.01$). Each model also
predicts $f_{\rho} \approx$ 1-2, implying that the two accretion rate
regimes would be active for similar fractions of time.

It is impossible with the current data to distinguish which, if either, of
these two explanations is correct for the evolution of the jet-mode
RLF. However, extending this analysis to still higher redshift would lead
to a clear distinction between the two. The purple dotted lines on the
middle and lower panels of Figure~\ref{fig_jetmodemodels} demonstrate the
predictions of Models 2b and 3a for the RLF of jet-mode radio-AGN at
redshift $z=1.5$, and these differ by an order of magnitude at most radio
luminosities. Extending the analysis of this paper to higher redshifts is
therefore critical, albeit that this will require near-infrared
spectroscopy if source classification is to be consistently carried out
using the oxygen lines.

\subsection{Evolution of the jet-mode AGN heating rate}

\begin{figure}
\begin{center}
\psfig{file=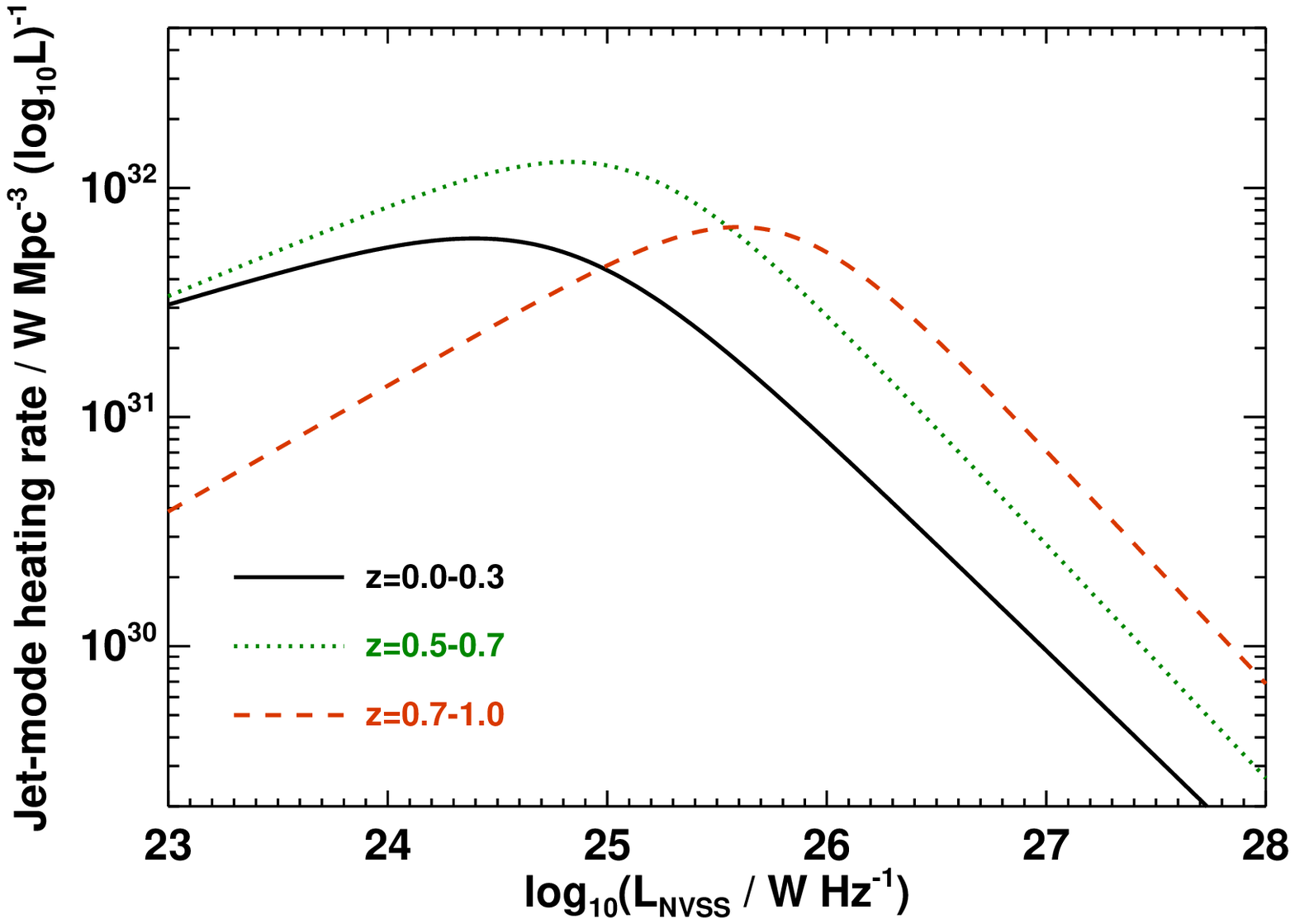,width=8.6cm,clip=}
\\
\psfig{file=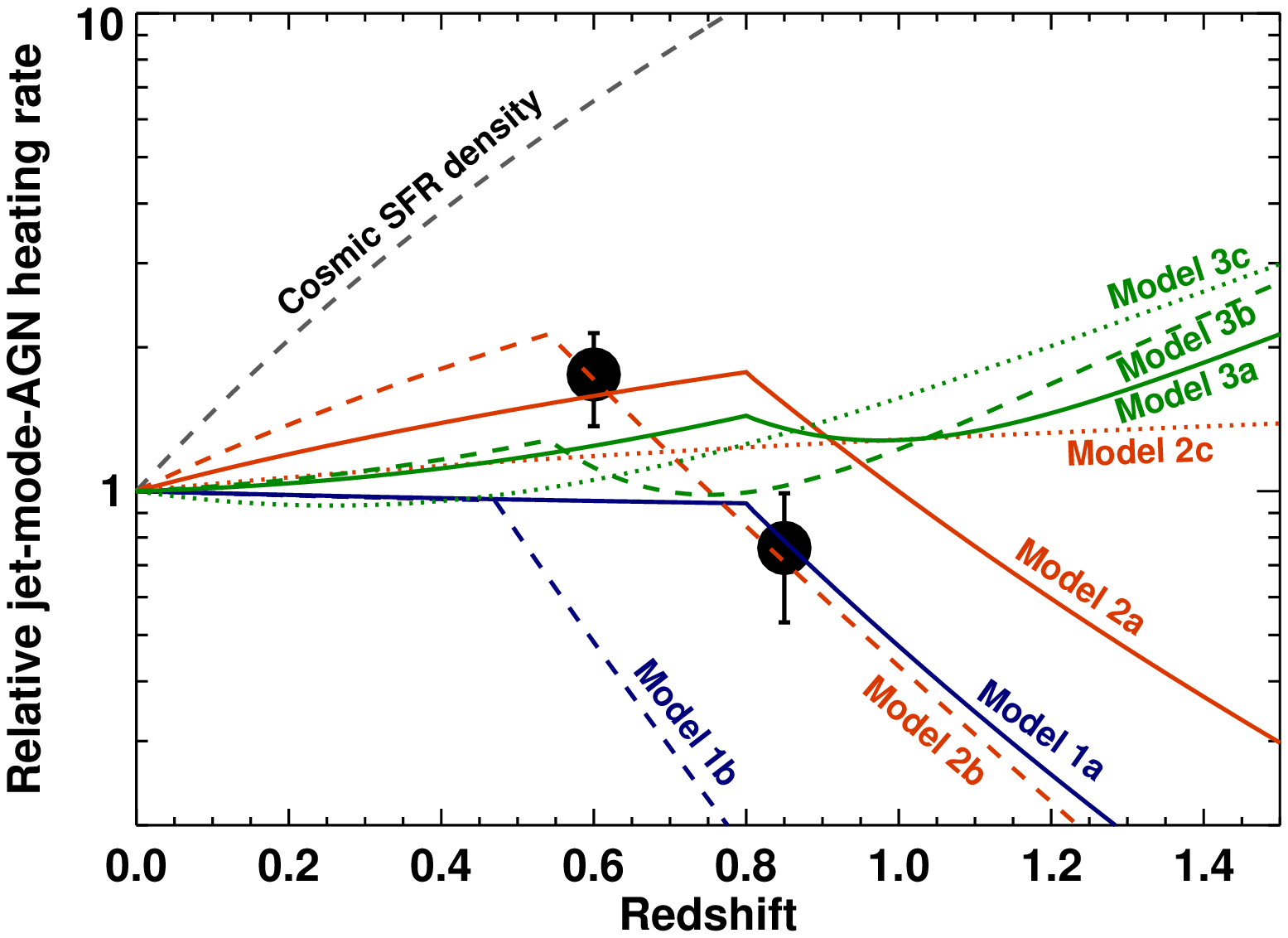,width=8.6cm,clip=}
\end{center}
\caption{\label{fig_heatrate} Top: the heating rate function of jet-mode
  radio-AGN over three redshift ranges, assuming that the relationship
  between radio luminosity and jet mechanical energy derived locally still
  holds at higher redshift. Bottom: the total integrated heating rate of
  jet-mode radio-AGN at redshifts $0.5 < z < 0.7$ and $0.7 < z < 1.0$
  compared to the local value. Also shown are the predictions for the
  evolution from various models described in Section~\ref{sec_jetmodels}.
  For comparison, the evolution of the cosmic star formation rate density
  is also shown, while the evolution of the space density of massive
  quiescent galaxies effectively follows Model 1a.}
\end{figure}

It is interesting to consider the implications of these results for the
importance of AGN-feedback as a function of redshift. As described by
\citet{hec14} and references therein, radio luminosity can be broadly
converted into a jet mechanical luminosity as 

\begin{displaymath}
P_{\rm mech} = 7 \times 10^{36} f_{\rm cav} \left(\frac{L_{\rm 1.4
    GHz}}{10^{25}{\rm W Hz}^{-1}}\right)^{0.68}\,{\rm W}
\end{displaymath}

\noindent where $f_{\rm cav} \approx 4$ relates the work done in inflating
the radio lobes to their pressure and volume, $E_{\rm cav} = f_{\rm cav}
pV$. If this relation remains invariant with redshift (which is not
necessarily the case if radio luminosities are boosted at higher redshifts
by higher confining gas densities) then by combining this relationship
with the RLF, the heating rate function as a function of radio luminosity
can be derived. This is shown for jet-mode radio-AGN in the top panel of
Figure~\ref{fig_heatrate}, as calculated from the broken power-law fits to
the RLF at each of the three redshifts. As can be seen, locally the
majority of heating arises from relatively low luminosity sources, but at
higher redshift the peak moves out to the higher luminosity
population. The overall heating rate per unit volume is found by
integrating this curve and is shown, relative to the local value, in the
lower panel of Figure~\ref{fig_heatrate}: it rises out to $z \sim 0.5$ but
then falls again. The predictions of the various models from
Section~\ref{sec_jetmodels} are also shown on this plot. For comparison,
the evolution of the cosmic star formation rate density (ie., broadly, the
cold gas supply) is also shown, while the evolution of the space density
of massive quiescent galaxies (ie., potential jet-mode AGN hosts) broadly
follows Model 1a (by definition of that model). Once again it can be seen
that Model 2b (luminosity-density evolution with a time-delay) provides
the best match to the data, and that the model predictions diverge
strongly towards higher redshifts.

These results can also be compared against previous phenomenological
models of jet-mode radio-AGN activity. \citet{cro06} predicted, in their
semi-analytic modelling incorporating radio-AGN feedback, that the
accretion rate onto radio AGN should remain relatively flat out to $z \sim
1.5$ and then fall by an order of magnitude out to $z \approx
4$. \citet{kor08} developed a model for the accretion rate function of
low-luminosity black holes based on the radio core emission, to predict
that the energetic output of jet-mode AGN should rise by a factor 2-3
between $z=0$ and $z=0.5$, and then remain broadly flat out to $z = 3$.
\citet{mer08} similarly used an AGN synthesis model to predict that the
kinetic energy output of jet-mode radio-AGN (their `LK' population) should
be flat or shallowly rising out to $z \approx 1$ and then gradually fall
thereafter. \citet{moc13} combined this AGN population model with a
parameterisation of the evolution of the accretion rates on to black
holes, to make a very similar prediction for the LK population. All of
these predictions are in broad agreement with the data, although none
match precisely. Higher accuracy determinations and an extension to
higher redshift would allow a more critical test of these models.

\section{Conclusions}
\label{sec_concs}

This paper presents the first observational measurement of the cosmic
evolution of jet-mode radio-AGN feedback out to $z=1$.  Eight flux-limited
radio source samples with high spectroscopic completeness are combined to
produce a catalogue of over 200 radio sources at redshifts $0.5 < z <
1.0$, which are then spectroscopically classified into jet-mode and
radiative-mode AGN classes. By comparison with the large samples of local
radio-AGN selected from the Sloan Digital Sky Survey \citep{bes12}, the
cosmic evolution of the RLF of each radio source
class is derived independently. 

Radiative-mode radio-AGN show a monotonic increase in space density with
redshift out to $z=1$, in line with the increasing space density of cosmic
star formation. This is consistent with these AGN being fuelled by cold
gas. Jet-mode radio-AGN show more complicated behaviour. At low radio
luminosities ($L_{\rm 1.4 GHz} \lta 10^{24}$W\,Hz$^{-1}$) their space
density decreases gradually with increasing redshift. At intermediate
luminosities ($10^{24}-10^{26}$W\,Hz$^{-1}$) it rises out to $z \approx
0.5$ and then falls at higher redshift. At the highest radio luminosities
the space density continues to increase out to $z=1$. 

Simple models are developed to explain the observed evolution. The
characteristic space density of jet-mode AGN is modelled as decreasing
with redshift in accordance with the number of massive quiescent galaxies,
in which they are believed to be hosted. A time delay between the
formation of the quiescent galaxy and its availability as a jet-mode
radio-AGN host is allowed, in case there is a lag between the quenching of
star formation activity and the onset of the hot gas cooling flows in the
galaxy which fuel the jet-mode AGN. The best-fitting models prefer a time
delay of 1.5--2\,Gyr which, intriguingly, is in line with the typical
cooling time of hot gas at radii of $\approx 10$\,kpc around massive
ellipticals.

The evolution at higher radio luminosities can be accounted for either by
allowing for evolution of the characteristic luminosity of the jet-mode
RLF with redshift (roughly as $(1+z)^3$) or if the jet-mode radio-AGN
population also includes some contribution of cold-gas-fuelled sources
hosted by the typical hosts of radiative-mode AGN, just caught at a time
when their accretion rate was low. The current data are unable to
distinguish between these two possibilities, but it is shown that
extending the analysis to still higher redshifts would provide a very
clear diagnostic.

If the relationship between jet mechanical luminosity and radio luminosity
remains constant across cosmic time then the results indicate that the
volume-averaged energetic output of jet-mode radio-AGN (ie. of radio-AGN
feedback) rises gradually out to about $z \sim 0.5$ and then falls beyond
that. This is broadly in line with the expectations of semi-analytic and
phenomenological models. Extending the analysis to higher redshifts and
improving the accuracy with larger radio source samples would provide a
more critical test of these models.

\section*{Acknowledgements} 

PNB, LMK and JS are grateful for financial support from STFC.  The William
Herschel Telescope is operated on the island of La Palma by the Isaac
Newton Group in the Spanish Observatorio del Roque de los Muchachos of the
Instituto de Astrof{\i}sica de Canarias. WHT data were obtained under
programmes W/12A/P16 and W/12B/P7. The authors thank Matt Jarvis for
helpful discussions.

\bibliography{pnb} 
\bibliographystyle{mn2e}

\appendix

\section{The radio source sample data}
\label{app_samples}

\begin{table*}
\caption{The properties of the 211-source combined sample used in the
  analysis of this paper. \label{tab_sample}}

\begin{tabular}{lclcrcl}
\hline
Source & S$_{\rm 1.4GHz}$ & ~Redshift  & Type & Spec. & log$_{10}$(L) & Classification \\      
       &  (Jy)           &            & of z & Index & (W\,Hz$^{-1}$) & \\
\hline
{\bf WP85:}~~~~~~~~~~~~~ \\
0538+49       & 21.79   &   0.545  & S &  0.77 &  28.38 &  Radiative-mode         \\   
1828+48       & 16.69   &   0.692  & S &  0.78 &  28.49 &  Quasar (radiative-mode)\\   
1328+30       & 14.70   &   0.850  & S &  0.53 &  28.59 &  Quasar (radiative-mode)\\   
0809+48       & 14.37   &   0.871  & S &  0.94 &  28.71 &  Quasar (radiative-mode)\\
0407-65       & 13.47   &   0.962  & S &  1.11 &  28.84 &  Radiative-mode         \\   
0518+16       & 12.99   &   0.759  & S &  0.92 &  28.52 &  Quasar (radiative-mode)\\   
0409-75       & 12.72   &   0.693  & S &  0.86 &  28.40 &  Radiative-mode         \\   
1458+71       & 8.89    &   0.905  & S &  0.77 &  28.49 &  Radiative-mode         \\   
0316+16       & 8.01    &   0.907  & S &  0.79 &  28.46 &  Radiative-mode         \\   
2032-35       & 7.62    &   0.631  & S &  1.10 &  28.13 &  Radiative-mode         \\   
1005+07       & 6.62    &   0.877  & S &  0.97 &  28.39 &  Radiative-mode         \\     
0252-71       & 6.55    &   0.568  & S &  1.14 &  27.96 &  Radiative-mode         \\   
1157+73       & 6.41    &   0.97   & S &  0.70 &  28.40 &  Radiative-mode         \\   
1609+66       & 6.19    &   0.550  & S &  0.76 &  27.83 &  Radiative-mode         \\   
0404+76       & 6.01    &   0.599  & S &  0.60 &  27.87 &  Radiative-mode         \\   
1254+47       & 5.59    &   0.996  & S &  1.02 &  28.47 &  Radiative-mode         \\
1634+62       & 5.09    &   0.988  & S &  0.96 &  28.40 &  Quasar (radiative-mode)\\   
1526-42       & 5.08    &   0.5    & P &  1.02 &  27.69 &  Unclassified           \\   
0117-15       & 4.91    &   0.565  & S &  0.90 &  27.78 &  Radiative-mode         \\   
2128+04       & 4.84    &   0.99   & S &  0.67 &  28.29 &  Unclassified           \\   
2331-41       & 4.84    &   0.907  & S &  0.91 &  28.27 &  Radiative-mode         \\   
0022-42       & 4.71    &   0.937  & S &  0.77 &  28.26 &  Radiative-mode         \\   
1453-10       & 4.6     &   0.938  & S &  0.93 &  28.29 &  Quasar (radiative-mode)\\   
1637+62       & 4.45    &   0.750  & S &  1.03 &  28.07 &  Radiative-mode         \\   
2342+82       & 4.35    &   0.735  & S &  0.95 &  28.01 &  Quasar (radiative-mode)\\   
1136-13       & 4.29    &   0.556  & S &  0.65 &  27.65 &  Quasar (radiative-mode)\\   
0235-19       & 4.27    &   0.620  & S &  0.87 &  27.81 &  Radiative-mode         \\   
2135-20       & 4.27    &   0.636  & S &  0.82 &  27.83 &  Radiative-mode         \\   
0157-31       & 4.03    &   0.677  & S &  0.81 &  27.87 &  Quasar (radiative-mode)\\   
\hline	               
{\bf CoNFIG:}\\
3C216         & 4.23    &   0.67   & S &  0.77 &  27.87 &  Quasar (radiative-mode)\\   
3C228         & 3.71    &   0.552  & S &  0.84 &  27.63 &  Radiative-mode         \\   
3C225         & 3.34    &   0.58   & S &  1.01 &  27.66 &  Radiative-mode         \\   
3C337         & 3.16    &   0.63   & S &  0.67 &  27.65 &  Radiative-mode         \\   
3C254         & 3.13    &   0.736  & S &  0.92 &  27.87 &  Quasar (radiative-mode)\\   
4C-06.35      & 2.96    &   0.625  & S &  0.88 &  27.66 &  Quasar (radiative-mode)\\   
3C275.1       & 2.90    &   0.557  & S &  0.91 &  27.54 &  Quasar (radiative-mode)\\   
3C265         & 2.89    &   0.811  & S &  0.94 &  27.94 &  Radiative-mode         \\   
3C247         & 2.88    &   0.749  & S &  0.84 &  27.83 &  Radiative-mode         \\   
4C03.18       & 2.71    &   0.535  & S &  0.52 &  27.40 &  Radiative-mode         \\   
3C207         & 2.61    &   0.680  & S &  0.81 &  27.68 &  Quasar (radiative-mode)\\   
3C336         & 2.61    &   0.927  & S &  0.73 &  27.98 &  Quasar (radiative-mode)\\   
3C340         & 2.60    &   0.775  & S &  0.68 &  27.78 &  Radiative-mode         \\   
4C19.44       & 2.59    &   0.72   & S &  0.53 &  27.67 &  Quasar (radiative-mode)\\   
4C33.21       & 2.47    &   0.603  & S &  0.56 &  27.48 &  Radiative-mode         \\   
4C01.39       & 2.40    &   0.819  & S &  0.70 &  27.81 &  Radiative-mode         \\   
3C289         & 2.40    &   0.967  & S &  0.79 &  28.00 &  Radiative-mode         \\   
3C226         & 2.39    &   0.818  & S &  0.93 &  27.86 &  Radiative-mode         \\   
4C01.42       & 2.26    &   0.792  & S &  0.66 &  27.74 &  Radiative-mode         \\   
4C37.24       & 2.26    &   0.919  & S &  0.65 &  27.88 &  Quasar (radiative-mode)\\   
4C59.16       & 2.18    &   0.961  & S &  0.56 &  27.88 &  Radiative-mode         \\   
3C217         & 2.09    &   0.898  & S &  0.95 &  27.91 &  Radiative-mode         \\   
3C334         & 1.99    &   0.555  & S &  0.84 &  27.36 &  Quasar (radiative-mode)\\   
3C277.2       & 1.95    &   0.766  & S &  0.89 &  27.70 &  Radiative-mode         \\   
1355+01       & 1.92    &   0.797  & S &  0.70 &  27.68 &  Radiative-mode         \\   
3C202         & 1.88    &   0.809  & S &  0.72 &  27.69 &  Radiative-mode         \\   
3C352         & 1.87    &   0.806  & S &  0.90 &  27.73 &  Radiative-mode         \\   
4C20.33       & 1.81    &   0.871  & S &  0.70 &  27.75 &  Quasar (radiative-mode)\\   
4C13.56       & 1.81    &   0.672  & S &  0.65 &  27.47 &  Radiative-mode         \\   
4C54.25       & 1.74    &   0.716  & S &  0.56 &  27.50 &  Unclassified           \\   
\hline
\end{tabular}
\end{table*}

\addtocounter{table}{-1}

\begin{table*}
\caption{continued}
\begin{tabular}{lclcrcl}
\hline
Source & S$_{\rm 1.4GHz}$ & ~Redshift  & Type & Spec. & log$_{10}$(L) & Classification \\      
       &  (Jy)           &            & of z & Index & (W\,Hz$^{-1}$) & \\
\hline
{\bf CoNFIG (cont.):}\\
4C53.18       & 1.60    &   0.869  & S &  0.77 &  27.71 &  Unclassified           \\   
4C17.56       & 1.57    &   0.777  & S &  0.62 &  27.55 &  Radiative-mode         \\   
4C43.22       & 1.57    &   0.572  & S &  0.70 &  27.26 &  Radiative-mode         \\   
4C24.31       & 1.56    &   0.653  & S &  0.78 &  27.41 &  Quasar (radiative-mode)\\   
4C17.48       & 1.53    &   0.521  & S &  0.74 &  27.16 &  Radiative-mode         \\   
4C04.40       & 1.50    &   0.531  & S &  0.83 &  27.19 &  Jet-mode               \\   
3C288.1       & 1.49    &   0.964  & S &  0.88 &  27.82 &  Quasar (radiative-mode)\\   
4C-00.50      & 1.47    &   0.892  & S &  0.56 &  27.64 &  Quasar (radiative-mode)\\   
4C46.21       & 1.44    &   0.527  & S &  0.72 &  27.14 &  Radiative-mode         \\   
3C344         & 1.42    &   0.52   & S &  0.88 &  27.15 &  Unclassified           \\   
4C61.34       & 1.35    &   0.523  & S &  0.67 &  27.10 &  Quasar (radiative-mode)\\   
3C272         & 1.35    &   0.944  & S &  0.98 &  27.78 &  Radiative-mode         \\   
3C323         & 1.34    &   0.679  & S &  0.91 &  27.41 &  Radiative-mode         \\   
3C342         & 1.34    &   0.561  & S &  0.81 &  27.19 &  Quasar (radiative-mode)\\   
4C51.25       & 1.31    &   0.561  & S &  0.81 &  27.18 &  Radiative-mode         \\   
4C20.29       & 1.27    &   0.68   & S &  0.66 &  27.33 &  Quasar (radiative-mode)\\   
4C32.34       & 1.26    &   0.564  & S &  0.98 &  27.21 &  Radiative-mode         \\   
3C232         & 1.25    &   0.531  & S &  0.79 &  27.10 &  Quasar (radiative-mode)\\   
4C46.25       & 1.16    &   0.743  & S &  0.71 &  27.39 &  Unclassified           \\   
3C261         & 1.15    &   0.613  & S &  1.00 &  27.26 &  Quasar (radiative-mode)\\   
3C281         & 1.12    &   0.599  & S &  0.87 &  27.20 &  Quasar (radiative-mode)\\   
4C59.11       & 1.08    &   0.707  & S &  0.88 &  27.35 &  Radiative-mode         \\   
4C00.35       & 1.08    &   0.746  & S &  0.71 &  27.37 &  Radiative-mode         \\   
4C15.34       & 1.07    &   0.97   & P &  0.72 &  27.64 &  Unclassified           \\   
4C17.49       & 1.06    &   0.51   & P &  0.90 &  27.01 &  Jet-mode               \\   
4C17.54       & 1.01    &   0.675  & S &  0.68 &  27.23 &  Radiative-mode         \\   
3C248         & 0.99    &   0.83   & P &  1.05 &  27.53 &  Unclassified           \\   
4C49.21       & 0.96    &   0.73   & P &  0.93 &  27.35 &  Unclassified           \\   
4C-02.43      & 0.96    &   0.70   & P &  0.81 &  27.28 &  Unclassified           \\   
4C00.34       & 0.92    &   0.906  & S &  0.94 &  27.56 &  Quasar (radiative-mode)\\   
3C251         & 0.92    &   0.781  & S &  0.89 &  27.39 &  Quasar (radiative-mode)\\   
1300+585      & 0.90    &   0.80   & P &  0.51 &  27.30 &  Unclassified           \\   
4C25.36       & 0.88    &   0.98   & P &  0.81 &  27.59 &  Unclassified           \\   
4C10.33       & 0.88    &   0.540  & S &  0.88 &  26.98 &  Quasar (radiative-mode)\\   
4C16.30       & 0.87    &   0.630  & S &  0.60 &  27.08 &  Quasar (radiative-mode)\\   
4C43.19       & 0.85    &   0.81   & P &  0.96 &  27.41 &  Jet-mode               \\   
4C09.39       & 0.81    &   0.696  & S &  0.59 &  27.14 &  Quasar (radiative-mode)\\   
1229-013      & 0.81    &   0.59   & P &  0.68 &  27.00 &  Unclassified           \\   
4C31.40       & 0.81    &   0.828  & S &  1.06 &  27.44 &  Radiative-mode         \\   
\hline		    
{\bf Parkes:}\\
2154-184      & 2.39    &   0.668  & S &  1.17 &  27.70 &  Quasar (radiative-mode)\\   
0222-008      & 1.11    &   0.687  & S &  0.79 &  27.31 &  Quasar (radiative-mode)\\   
2355-010      & 0.83    &   0.76   & P &  1.02 &  27.35 &  Unclassified           \\   
0059+017      & 0.80    &   0.692  & S &  1.06 &  27.24 &  Radiative-mode         \\   
1336+020      & 0.74    &   0.567  & S &  1.02 &  26.99 &  Radiative-mode         \\   
2159-201      & 0.60    &   0.75   & P &  1.66 &  27.35 &  Unclassified           \\   
2158-177      & 0.54    &   0.81   & P &  0.93 &  27.21 &  Unclassified           \\   
0242+028      & 0.53    &   0.767  & S &  0.95 &  27.15 &  Radiative-mode         \\   
0043+000      & 0.53    &   0.60   & P &  1.03 &  26.90 &  Unclassified           \\   
\hline
{\bf 7CRS:}\\
1732+6535     & 0.73    &   0.856  & S &  0.98 &  27.41 &  Quasar (radiative-mode)\\   
5C6.19        & 0.44    &   0.799  & S &  0.80 &  27.07 &  Radiative-mode         \\   
0221+3417     & 0.44    &   0.852  & S &  0.87 &  27.15 &  Unclassified           \\   
1816+6710     & 0.40    &   0.92   & S &  0.95 &  27.21 &  Jet-mode               \\   
5C7.111       & 0.37    &   0.628  & S &  0.77 &  26.74 &  Radiative-mode         \\   
1807+6831     & 0.37    &   0.58   & S &  0.79 &  26.66 &  Radiative-mode         \\   
5C7.8         & 0.30    &   0.673  & S &  0.89 &  26.75 &  Radiative-mode         \\   
1755+6830     & 0.28    &   0.744  & S &  0.85 &  26.81 &  Radiative-mode         \\   
5C7.85        & 0.27    &   0.995  & S &  0.78 &  27.08 &  Quasar (radiative-mode)\\   
5C7.205       & 0.26    &   0.710  & S &  0.88 &  26.74 &  Radiative-mode         \\   
\hline
\end{tabular}
\end{table*}
\addtocounter{table}{-1}

\begin{table*}
\caption{continued}
\begin{tabular}{lclcrcl}
\hline
Source & S$_{\rm 1.4GHz}$ & ~Redshift  & Type & Spec. & log$_{10}$(L) & Classification \\      
       &  (Jy)           &            & of z & Index & (W\,Hz$^{-1}$) & \\
\hline
{\bf 7CRS (cont.):}\\
5C7.118       & 0.25    &   0.527  & S &  0.71 &  26.39 &  Radiative-mode         \\   
1819+6550     & 0.23    &   0.724  & S &  0.83 &  26.70 &  Quasar (radiative-mode)\\   
1758+6535     & 0.23    &   0.80   & S &  0.78 &  26.79 &  Radiative-mode         \\   
1826+6510     & 0.22    &   0.646  & S &  0.94 &  26.59 &  Jet-mode               \\   
1815+6815     & 0.19    &   0.794  & S &  1.03 &  26.77 &  Jet-mode               \\   
5C7.25        & 0.19    &   0.671  & S &  0.66 &  26.50 &  Radiative-mode         \\   
5C6.264       & 0.19    &   0.831  & S &  0.82 &  26.75 &  Quasar (radiative-mode)\\   
5C6.258       & 0.19    &   0.752  & S &  0.61 &  26.59 &  Unclassified           \\   
5C7.125       & 0.19    &   0.801  & S &  0.63 &  26.65 &  Radiative-mode         \\   
1816+6605     & 0.18    &   0.92   & S &  0.95 &  26.87 &  Unclassified           \\   
5C6.233       & 0.18    &   0.560  & S &  0.95 &  26.34 &  Radiative-mode         \\   
\hline		      	       	                             
{\bf TOOT:}\\
TOOT00\_1200  & 0.27    &   0.691  & S &  0.79 &  26.71 &  Radiative-mode         \\   
TOOT00\_1034  & 0.14    &   0.580  & S &  0.66 &  26.22 &  Jet-mode               \\   
TOOT00\_1140  & 0.12    &   0.911  & S &  0.66 &  26.61 &  Jet-mode               \\   
TOOT00\_1072  & 0.11    &   0.577  & S &  0.72 &  26.12 &  Unclassified           \\   
TOOT00\_1267  & 0.100   &   0.968  & S &  0.74 &  26.61 &  Radiative-mode         \\   
TOOT00\_1235  & 0.099   &   0.743  & S &  0.55 &  26.29 &  Quasar (radiative-mode)\\   
TOOT00\_1255  & 0.034   &   0.582  & S &  0.60 &  25.59 &  Unclassified           \\   
\hline		      	       	                             
{\bf CENSORS:}~~~~~~~~~~ \\
CEN-6         & 0.24    &   0.547  & S &  0.54 &  26.37 &  Radiative-mode         \\   
CEN-12        & 0.070   &   0.821  & S &  0.95 &  26.34 &  Radiative-mode         \\   
CEN-17        & 0.062   &   0.893  & S &  0.73 &  26.31 &  Unclassified           \\   
CEN-22        & 0.053   &   0.928  & S &  0.95 &  26.35 &  Radiative-mode         \\   
CEN-29        & 0.038   &   0.965  & S &  0.80 &  26.20 &  Radiative-mode         \\   
CEN-37        & 0.032   &   0.511  & S &  0.76 &  25.47 &  Unclassified           \\   
CEN-43        & 0.026   &   0.778  & S &  0.65 &  25.78 &  Unclassified           \\   
CEN-45        & 0.026   &   0.796  & S &  0.89 &  25.85 &  Jet-mode               \\   
CEN-47        & 0.025   &   0.508  & S &  0.81 &  25.37 &  Radiative-mode         \\   
CEN-55        & 0.021   &   0.557  & S &  0.65 &  25.36 &  Jet-mode               \\   
CEN-62        & 0.018   &   0.574  & S &  0.63 &  25.32 &  Jet-mode               \\   
CEN-65        & 0.018   &   0.549  & S &  0.58 &  25.25 &  Jet-mode               \\   
CEN-70        & 0.017   &   0.645  & S &  0.94 &  25.47 &  Radiative-mode         \\   
CEN-74        & 0.016   &   0.667  & S &  0.92 &  25.47 &  Radiative-mode         \\   
CEN-83        & 0.014   &   0.521  & S &  0.98 &  25.15 &  Radiative-mode         \\   
CEN-86        & 0.013   &   0.82   & P &  0.62 &  25.53 &  Unclassified           \\   
CEN-89        & 0.013   &   0.909  & S &  1.00 &  25.73 &  Unclassified           \\   
CEN-92        & 0.013   &   0.743  & S &  1.27 &  25.57 &  Radiative-mode         \\   
CEN-104       & 0.011   &   0.88   & P &  1.19 &  25.67 &  Jet-mode               \\   
CEN-107       & 0.010   &   0.512  & S &  0.82 &  24.99 &  Jet-mode               \\   
CEN-109       & 0.010   &   0.72   & P &  0.91 &  25.35 &  Jet-mode               \\   
CEN-113       & 0.0097  &   0.94   & P &  1.23 &  25.71 &  Jet-mode               \\   
CEN-115       & 0.0096  &   0.545  & S &  1.20 &  25.09 &  Jet-mode               \\   
CEN-125       & 0.0084  &   0.701  & S &  1.02 &  25.27 &  Jet-mode               \\   
CEN-127       & 0.0083  &   0.922  & S &  1.23 &  25.61 &  Unclassified           \\   
CEN-136       & 0.0075  &   0.629  & S &  1.08 &  25.11 &  Unclassified           \\   
CEN-137       & 0.0074  &   0.526  & S &  0.93 &  24.89 &  Jet-mode               \\   
CEN-138       & 0.015   &   0.508  & S &  0.71 &  25.11 &  Unclassified           \\   
\hline
{\bf Hercules:}~~~~~~~~~ \\	     	       	                             
53W008        & 0.31    &   0.736  & S &  0.79 &  26.83 &  Radiative-mode         \\   
53W031        & 0.12    &   0.627  & S &  0.70 &  26.22 &  Radiative-mode         \\   
53W023        & 0.11    &   0.569  & S &  0.87 &  26.13 &  Jet-mode               \\   
53W046        & 0.063   &   0.528  & S &  0.69 &  25.78 &  Radiative-mode         \\   
53W080        & 0.028   &   0.542  & S &  0.80 &  25.47 &  Quasar (radiative-mode)\\   
53W047        & 0.024   &   0.532  & S &  0.67 &  25.36 &  Jet-mode               \\   
53W067        & 0.023   &   0.759  & S &  0.81 &  25.74 &  Jet-mode               \\   
53W026        & 0.021   &   0.550  & S &  0.74 &  25.36 &  Jet-mode               \\   
53W048        & 0.012   &   0.676  & S &  0.81 &  25.32 &  Jet-mode               \\   
53W060        & 0.010   &   0.62   & P &  0.93 &  25.18 &  Unclassified           \\   
\hline
\end{tabular}
\end{table*}

\addtocounter{table}{-1}

\begin{table*}
\caption{continued}
\begin{tabular}{lclcrcl}
\hline
Source & S$_{\rm 1.4GHz}$ & ~Redshift  & Type & Spec. & log$_{10}$(L) & Classification \\      
       &  (Jy)           &            & of z & Index & (W\,Hz$^{-1}$) & \\
\hline
{\bf Hercules (cont.):}~~~~ \\	     	       	                             
53W041        & 0.0094  &   0.59   & P &  0.88 &  25.10 &  Unclassified           \\   
53W077        & 0.0078  &   0.786  & S &  0.87 &  25.32 &  Jet-mode               \\   
53W005        & 0.0076  &   0.765  & S &  1.09 &  25.33 &  Unclassified           \\   
53W019        & 0.0068  &   0.542  & S &  0.72 &  24.85 &  Jet-mode               \\   
53W083        & 0.0050  &   0.628  & S &  0.70 &  24.86 &  Jet-mode               \\   
53W089        & 0.0025  &   0.635  & S &  1.29 &  24.69 &  Jet-mode               \\   
\hline		     
{\bf SXDF:} \\
VLA0001       & 0.080   &   0.627  & S &  ---  &  26.08 &  Radiative-mode         \\   
VLA0011       & 0.0080  &   0.645  & S &  ---  &  25.11 &  Jet-mode               \\   
VLA0012       & 0.0066  &   0.865  & S &  ---  &  25.33 &  Unclassified           \\   
VLA0018       & 0.0048  &   0.919  & S &  ---  &  25.26 &  Jet-mode               \\
VLA0019       & 0.0048  &   0.695  & S &  ---  &  24.97 &  Unclassified           \\   
VLA0023       & 0.0042  &   0.586  & S &  ---  &  24.73 &  Radiative-mode         \\   
VLA0024       & 0.0036  &   0.516  & S &  ---  &  24.54 &  Jet-mode               \\   
VLA0025       & 0.0032  &   0.963  & S &  ---  &  25.13 &  Quasar (radiative-mode)\\   
VLA0028       & 0.0028  &   0.632  & S &  ---  &  24.63 &  Jet-mode               \\   
VLA0030       & 0.0026  &   0.535  & S &  ---  &  24.42 &  Jet-mode               \\   
VLA0033       & 0.0024  &   0.647  & S &  ---  &  24.58 &  Jet-mode               \\   
VLA0036       & 0.0021  &   0.872  & S &  ---  &  24.84 &  Jet-mode               \\   
VLA0039       & 0.0018  &   0.89   & P &  ---  &  24.78 &  Jet-mode               \\   
VLA0045       & 0.0015  &   0.553  & S &  ---  &  24.23 &  Jet-mode               \\   
VLA0060       & 0.0010  &   0.881  & S &  ---  &  24.53 &  Unclassified           \\   
VLA0061       & 0.00099 &   0.668  & S &  ---  &  24.24 &  Radiative-mode         \\   
VLA0064       & 0.00097 &   0.515  & S &  ---  &  23.97 &  Unclassified           \\   
VLA0067       & 0.00085 &   0.649  & S &  ---  &  24.14 &  Jet-mode               \\   
VLA0080       & 0.00072 &   0.99   & P &  ---  &  24.46 &  Jet-mode               \\   
VLA0083       & 0.00071 &   0.94   & P &  ---  &  24.47 &  Unclassified           \\   
VLA0108       & 0.00050 &   0.627  & S &  ---  &  23.88 &  Jet-mode               \\   
VLA0115       & 0.00047 &   0.88   & P &  ---  &  24.20 &  Unclassified           \\   
VLA0120       & 0.00044 &   0.843  & S &  ---  &  24.13 &  Radiative-mode         \\   
VLA0145       & 0.00035 &   0.547  & S &  ---  &  23.58 &  Jet-mode               \\   
VLA0148       & 0.00034 &   0.64   & P &  ---  &  23.73 &  Unclassified           \\   
VLA0151       & 0.00034 &   0.579  & S &  ---  &  23.63 &  Jet-mode               \\   
VLA0169       & 0.00030 &   0.553  & S &  ---  &  23.53 &  Jet-mode               \\   
VLA0177       & 0.00029 &   0.515  & S &  ---  &  23.44 &  Jet-mode               \\
VLA0186       & 0.00028 &   0.567  & S &  ---  &  23.51 &  Jet-mode               \\
VLA0202       & 0.00025 &   0.83   & P &  ---  &  23.87 &  Unclassified           \\ 
VLA0209       & 0.00025 &   0.76   & P &  ---  &  23.75 &  Unclassified           \\
VLA0217       & 0.00024 &   0.643  & S &  ---  &  23.57 &  Jet-mode               \\  
\hline		    	       		    		 
\end{tabular}	    
\end{table*}

Table~\ref{tab_sample} provides data for the sources within the eight
samples used for this study. The 1.4\,GHz flux density limits and sky area
coverage of each sample are given in Table~\ref{tab_surveys}. Redshifts are
selected to be between 0.5 and 1.0; where a spectroscopic redshift does
not exist then a photometric redshift (in some cases based only on the
K$-z$ relation) is used instead. These are indicated in the table. Readers
should refer back to the papers originally presenting the samples (as
described in Section~\ref{sec_samples}) for details of the origins of the
redshifts and other source properties. Spectral indices are calculated at
1.4\,GHz, following \citet{ker12}, although the effect of any spectral
curvature on the calculated radio luminosities is negligible at these
redshifts. No spectral indices are available for the SXDF sources, so a
value of $\alpha = 0.75$ is assumed. Classifications into jet-mode and
radiative-mode are described in Section~\ref{sec_classify}.

\section{New observational results}
\label{app_specdetails}

Spectroscopic observations were carried out of a subsample of the galaxies
which either lacked a spectroscopic redshift or for which no suitable
spectrum was available for classification. These observations were carried
out during two runs on the William Herschel Telescope (WHT) from 22-24 May
and 18-19 October 2012 (the latter of which was almost entirely lost to
bad weather), with further observations obtained in service mode in
November 2012. Observations were carried out using the duel--beam ISIS
spectrograph, with the 5300\AA\ dichroic, the R300B and R158R gratings in
the blue and red arms respectively, and a 1.5 arcsec slit. Combined, these
provided a usable wavelength coverage from $\approx 3500$\AA\ to $\approx
9500$\AA, and a spectral resolution of about 15\AA. 

Target exposure times varied according to the flux expected for emission
lines if the source was of radiative-mode class (for which emission line
flux correlates broadly with radio flux density). Initial integrations were
analysed in real-time (except for the service mode observations), and
observations were repeated if no classification was available, up to a
maximum of 40 mins. Data reduction was carried out in IRAF using standard
procedures, with internal calibration lamps used for flat-fielding and
wavelength calibration. Spectrophotometric calibration was carried out
using the standard star HZ21 for the May run and G191-B2B in October and
November.

\begin{table*}
\caption{\label{tab_included} Redshifts and emission line properties of 
sources in the sample which were observed in the new WHT observations.}
\begin{tabular}{ccccccccc}
\hline
Source & Observation & Exposure & Redshift & $f_{\rm [OII]}$ & $f_{\rm [OIII]}$ & $EW_{\rm [OII]}$& $EW_{\rm [OIII]}$ & Class \\
       &    date     & time / s & & \multicolumn{2}{c}{[$10^{-19}$W\,m$^{-2}$]} &\AA   &  \AA &       \\
\hline
4C33.21 &  2012-10-19 &  540 & 0.603 & 26.1 & 53   & 95 & 125 & Radiative-mode \\ 
4C59.16 &  2012-05-24 & 1200 & 0.961 & 13.4 & 13.1 & 22 & 22  & Radiative-mode \\
1355+01 &  2012-05-23 & 1200 & 0.797 & 35.0 & 90   & 63 & 88  & Radiative-mode \\  
3C202   &  2012-10-19 &  420 & 0.809 & 16.2 & 95   & 80 & 250 & Radiative-mode \\
4C13.56 &  2012-05-24 & 1200 & 0.672 & 4.7  &      & 16 &     & Unclassified   \\
4C54.25 &  2012-05-23 & 1200 & 0.716 & 2.9  &      & 11 &     & Unclassified   \\  
4C53.18 &  2012-05-22 & 1200 & 0.869 & 1.0  &      &  6 &     & Unclassified   \\
4C17.56 &  2012-05-24 & 1800 & 0.777 & 4.1  & 4.5  & 21 & 10  & Radiative-mode \\
4C17.48 &  2012-05-24 & 1200 & 0.521 & 4.9  & 25.0 & 16 & 48  & Radiative-mode \\  
4C04.40 &  2012-05-24 & 1200 & 0.531 &      &      &    &     & Jet-mode       \\
4C46.21 &  2012-05-23 & 1200 & 0.527 & 38.8 & 68.8 & 45 & 44  & Radiative-mode \\ 
4C51.25 &  2012-05-23 & 1200 & 0.561 & 5.4  & 22.3 & 10 & 18  & Radiative-mode \\ 
4C32.34 &  2012-05-23 & 1200 & 0.564 & 59.1 & 402  & 80 & 275 & Radiative-mode \\
4C59.11 &  2012-05-23 & 1200 & 0.707 &      & 72.4 &    & 15  & Radiative-mode \\       
4C00.35 &  2012-05-23 & 1200 & 0.746 & 13.1 & 12.6 & 19 & 10  & Radiative-mode \\       
4C17.54 &  2012-05-24 & 1200 & 0.675 & 14.1 & 14.5 & 80 & 44  & Radiative-mode \\
4C43.19 &  2012-10-19 & 1200 &       &      &      &    &     & Jet-mode       \\
PKS0059+017&2012-11-10&  900 & 0.692 & 6.5  & 10.6 & 250& 200 & Radiative-mode \\
PKS1336+020&2012-05-23& 1200 & 0.567 & 32.3 & 336  & 28 & 166 & Radiative-mode \\   
53W008  &  2012-05-22 & 2400 & 0.736 & 3.2  & 8.8  &  7 &  13 & Radiative-mode \\
53W031  &  2012-05-22 & 2400 & 0.627 & 0.97 & 1.65 &  8 &  5  & Radiative-mode \\
53W023  &  2012-05-22 & 2400 & 0.569 &      &      &    &     & Jet-mode       \\
53W080  &  2012-05-22 & 2400 & 0.542 & 12.3 & 36.1 &  7 & 23  & Radiative-mode \\
53W047  &  2012-05-22 & 2400 & 0.532 &      &      &    &     & Jet-mode       \\     
53W077  &  2012-05-23 & 2400 & 0.786 &      &      &    &     & Jet-mode       \\
53W005  &  2012-05-23 & 2400 & 0.765 & 3.7  &      & 13 &     & Unclassified   \\
\hline
\end{tabular}
\end{table*}

\begin{table*}
\caption{\label{tab_excluded} Redshifts and emission line properties for
  sources observed with the WHT that were initially within the sample
  based on their photometric redshifts, but excluded from analysis on the
  basis of their new spectroscopic redshifts (or, for PKS1329+012 and
  PKS0045-009, due to the final flux density limit applied to the Parkes
  sample).}
\begin{tabular}{cccccccc}
\hline
Source & Observation & Exposure & Redshift & $f_{\rm [OII]}$ & $f_{\rm [OIII]}$ & $EW_{\rm [OII]}$& $EW_{\rm [OIII]}$ \\
       &    date     & time / s & &\multicolumn{2}{c}{[$10^{-19}$W\,m$^{-2}$]} &\AA   &  \AA \\
\hline
4C16.27 &  2012-05-24 & 1200 & 1.452 & 3.4 &      & 10 &     \\
1152+551&  2012-05-23 & 1800 & 1.195 & 8.1 &      & 10 &     \\
4C29.46 &  2012-05-24 & 1200 & 0.397 &     & 177  &    & 132 \\ 
4C12.41 &  2012-05-24 & 1200 & 1.10  & 10.0&      & 16 &     \\
4C59.10 &  2012-05-22 & 1200 & 1.245 & 5.3 &      & 45  &    \\
4C20.28 &  2012-05-22 & 1200 & 0.424 & 8.2 & 53.1 & 42 & 115 \\
PKS1352+008&2012-05-22& 1200 & 1.167 & 1.7 &      & 12 &     \\
PKS1337-033&2012-05-24& 1200 & 0.487 & 1.6 & 4.9  &    & 129 \\
PKS1329+012&2012-05-24& 2400 & 0.873 & 1.1 &      & 10 &     \\
PKS0045-009&2012-11-10&  900 & 0.832 & 3.9 &      & 120&     \\
\hline  
\end{tabular}
\end{table*}

The properties of the resulting spectra are presented in
Table~\ref{tab_included} for observed sources that remained within the
final sample (equivalent widths are rest-frame values).
Table~\ref{tab_excluded} presents the data for sources that were excluded
from the final analysis because their spectroscopic redshift lay outside of
the range $0.5 < z < 1.0$ studied here. In addition to these, the source
PKS0010+005 was excluded from analysis based on the new WHT
observations. This source has a photometric redshift of 0.4, but a
spectroscopic redshift of 0.606 is quoted in \citet{dun89b} based on a
private communication from Spinrad. However, Spinrad's spectrum remains
unpublished and the WHT data failed to detect emission lines consistent
with this redshift (or to determine any other redshift). Therefore the
photometric redshift was adopted for PKS0010+005, leading to its exclusion
from the sample. Finally, Table~\ref{tab_excluded} includes data for
PKS1329+012 and PKS0045-009 which were observed with the WHT but fell
below the eventual flux density limit applied to the Parkes subsample.

\label{lastpage}
\end{document}